\documentclass[preprint,12pt]{elsarticle}
\biboptions{sort&compress}

\usepackage{verbatim} 
\usepackage{epsfig}
\usepackage{amssymb} 
\usepackage{mathtools}
\usepackage{lineno}
\usepackage{tikz}        
\usepackage{etoolbox}
\usepackage{ulem}         
\usepackage[caption=false]{subfig}
\usepackage{textcomp}

\begin{document}

\newcommand*{\hwplotB}{\raisebox{3pt}{\tikz{\draw[red,dashed,line 
width=3.2pt](0,0) -- 
(5mm,0);}}}

\newrobustcmd*{\mydiamond}[1]{\tikz{\filldraw[black,fill=#1] (0.025,0) -- 
(0.1cm,0.15cm) -- (0.2cm,0) -- (0.1cm,-0.15cm) -- (0.025,0);}}

\newrobustcmd*{\mytriangleright}[1]{\tikz{\filldraw[black,fill=#1] (0,0.15cm) 
-- (0.3cm,0) -- (0,-0.15cm) -- (0,0.15cm);}}

\newrobustcmd*{\mytriangleup}[1]{\tikz{\filldraw[black,fill=#1] (0,0.3cm) 
-- (0.2cm,0) -- (-0.2cm,0) -- (0,0.3cm);}}

\newrobustcmd*{\mytriangleleft}[1]{\tikz{\filldraw[black,fill=#1] (0,0.15cm) -- 
(-0.3cm,0) -- (0,-0.15cm) -- (0,0.15cm);}}
\definecolor{Blue}{cmyk}{1.,1.,0,0} 

\begin{frontmatter}



\title{COVID-19 spreading under containment actions}

\author[add1]{F.E.~Cornes}
 \address[add1]{Departamento de F\'\i sica, Facultad de Ciencias 
Exactas y Naturales, \\ Universidad de Buenos Aires,\\
 Pabell\'on I, Ciudad Universitaria, 1428 Buenos Aires, Argentina.}

 \author[add2]{G.A.~Frank}
 \address[add2]{Unidad de Investigaci\'on y Desarrollo de las 
Ingenier\'\i as, Universidad Tecnol\'ogica Nacional, Facultad Regional Buenos 
Aires, Av. Medrano 951, 1179 Buenos Aires, Argentina.}

\author[add1,add3]{C.O.~Dorso\corref{cor1}}%

 \address[add3]{Instituto de F\'\i sica de Buenos Aires,\\
Pabell\'on I, Ciudad Universitaria, 1428 Buenos Aires, Argentina.}

\begin{abstract}

We propose an epidemiological model that includes the mobility patterns of the 
individuals, in the spirit to those considered in 
Refs.~\cite{Barmak,Barmak2,Medus}. We assume that people move around in a city 
of 120$\times$120 blocks with 300 inhabitants in each block. The mobility 
pattern is associated to a complex network in which nodes represent blocks while 
the links represent the traveling path of the individuals (see below). We 
implemented three confinement strategies in order to mitigate the disease 
spreading: 1) global confinement, 2) partial restriction to mobility, and 3) 
localized confinement. In the first case, it was observed that 
a global isolation policy prevents the massive outbreak of the disease. In the 
second case, a partial restriction to mobility could lead to a massive 
contagion if this was not complemented with sanitary measures such as the 
use of masks and social distancing. Finally, a local isolation policy was 
proposed, conditioned to the health status of each block. It was observed 
that this mitigation strategy was able to contain and even reduce the outbreak 
of the disease by intervening in specific regions of the city according to their 
level of contagion. It was also observed that this strategy is capable of 
controlling the epidemic in the case that a certain proportion of those infected 
are asymptomatic.

\end{abstract}


\end{frontmatter}


\section{\label{sec:introduction}Introduction}

In the absence of a vaccine, strategies based on non-pharmaceutical 
interventions were proposed to contain the COVID-19 pandemic. Social distancing 
policies, specifically mobility restrictions and lockdowns, among others 
were the more common ones. Such policies should be implemented for long periods 
(typically months) to avoid re-emergence of the epidemic once lifted. 
Therefore, quantitative research is still needed to assess the efficacy of 
non-pharmaceutical interventions and their timings.\\

Many works analyze real-time mobility data in order to relate the changes in 
the mobility patterns and the disease propagation. Ref.~\cite{Badr} reports a 
correlation between the mobility pattern and the reduction of new 
infections. Besides, they found that it takes two to three weeks to see results 
due to the incubation time of the disease. Also, Ref.~\cite{Tian} carries out a 
detailed study of the effects of containment measures during the first 50 days 
of the COVID-19 epidemic in China. These researchers found that 
traveling restrictions and social distancing measures (among others) were 
effective in the containment of the disease. Ref.~\cite{Chang} analyzes real 
mobility datasets in many US metropolitan areas. They found that a small 
minority of ``super-spreaders'' places are the responsible for the wide 
propagation of the infection.\\

The changes in the mobility patterns are a consequence of the implementation of 
different quarantines. Ref.~\cite{Meidan} analyzes different 
quarantine types considering a complex SEIR scheme. They suggested an 
alternative type of quarantine to reduce the infection disease while allowing a 
socio-economic activity. In a similar way, Ref.~\cite{Alon} proposes a cyclic 
schedule of 4-days work and 10-days lockdown. Also, an improved version of this 
strategy can be found in Ref.~\cite{cornes2020}. The influence of human 
behaviors on infectious disease transmission \cite{Zhao}, the effects of 
vaccination during a pandemic \cite{Paltiel} and the role of the 
``super-spreaders'' \cite{Cave,Nielsen} are many other complex scenarios 
analyzed in the literature. However, a compressive and quantitative comparison 
of the effectivenesses of different lockdown and their timing appears to be 
lacking. \\

In this work we consider the spread of a disease mimicking the COVID-19, 
assuming a spatio-temporal SEIR model of mobile agents. We simulated 
and compared different confinement strategies in order to mitigate the disease 
propagation. In Section~\ref{sec:epidemiological_model} we describe the 
characteristics of the epidemiological model and the different mitigation 
strategies. Section~\ref{sec:simulations} details the simulation 
procedure. Section~\ref{sec:results} displays the results of our investigation, 
while Section~\ref{sec:discussion} is dedicated to the discussion of the main 
results. The conclusions are drawn in Section~\ref{sec:conclusions}.\\

\section{\label{sec:epidemiological_model}The epidemiological model}

There are three main ingredients in the description of the COVID-19 
contagion and spatial spreading: the scenario where the process takes place 
(Section~\ref{sec:scenario}), the mobility patterns of the individuals 
(Section~\ref{sec:mobility_pattern}) and the epidemiological dynamics of the 
individuals (Section~\ref{sec:seir}). \\

\subsection{\label{sec:scenario}The stage}

\subsection*{\label{sec:city}City}

The simulation of the evolution of COVID-19 is performed in a schematic city in 
which the basic unit is the block. The city is represented by an homogeneous 
urban area of $120\times120~$blocks placed in a square grid. The size of each 
block is $100\,$m$\times100\,$m. The simulated grid corresponds to a big 
city, like Buenos Aires, Argentina. Each block hosts 300 people but this 
quantity may vary during the day (see next Section). The total population of the 
simulated city is $4.32\,$M similar to the Ciudad Autonoma de Buenos Aires' 
population. \\

\subsection*{\label{sec:simulations_network}Network construction}

The human mobility pattern between the blocks is accomplished by building a 
weighted and directed network. The nodes represent the blocks, while the 
links represent the human mobility from one block to another one. We consider 
two different links: short and long links. \\

\begin{figure*}[!ht]
\subfloat[\scriptsize{Network with only short links}\label{fig:red_cortos}]{
\includegraphics[scale=0.3]{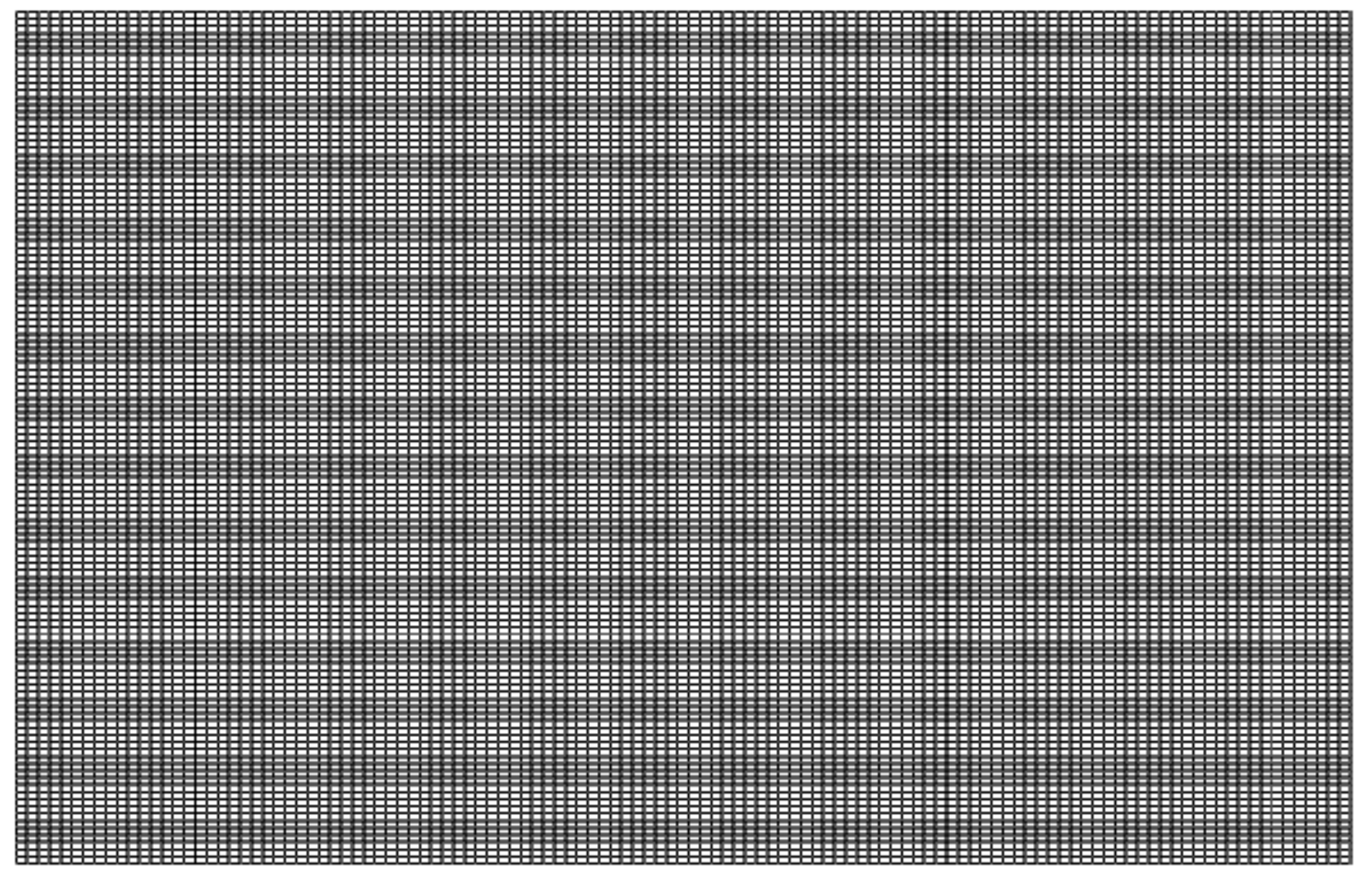}
}
\hspace{5mm}
\subfloat[\scriptsize{Network with only long links}\label{fig:red_largos}]{
\includegraphics[scale=0.3]{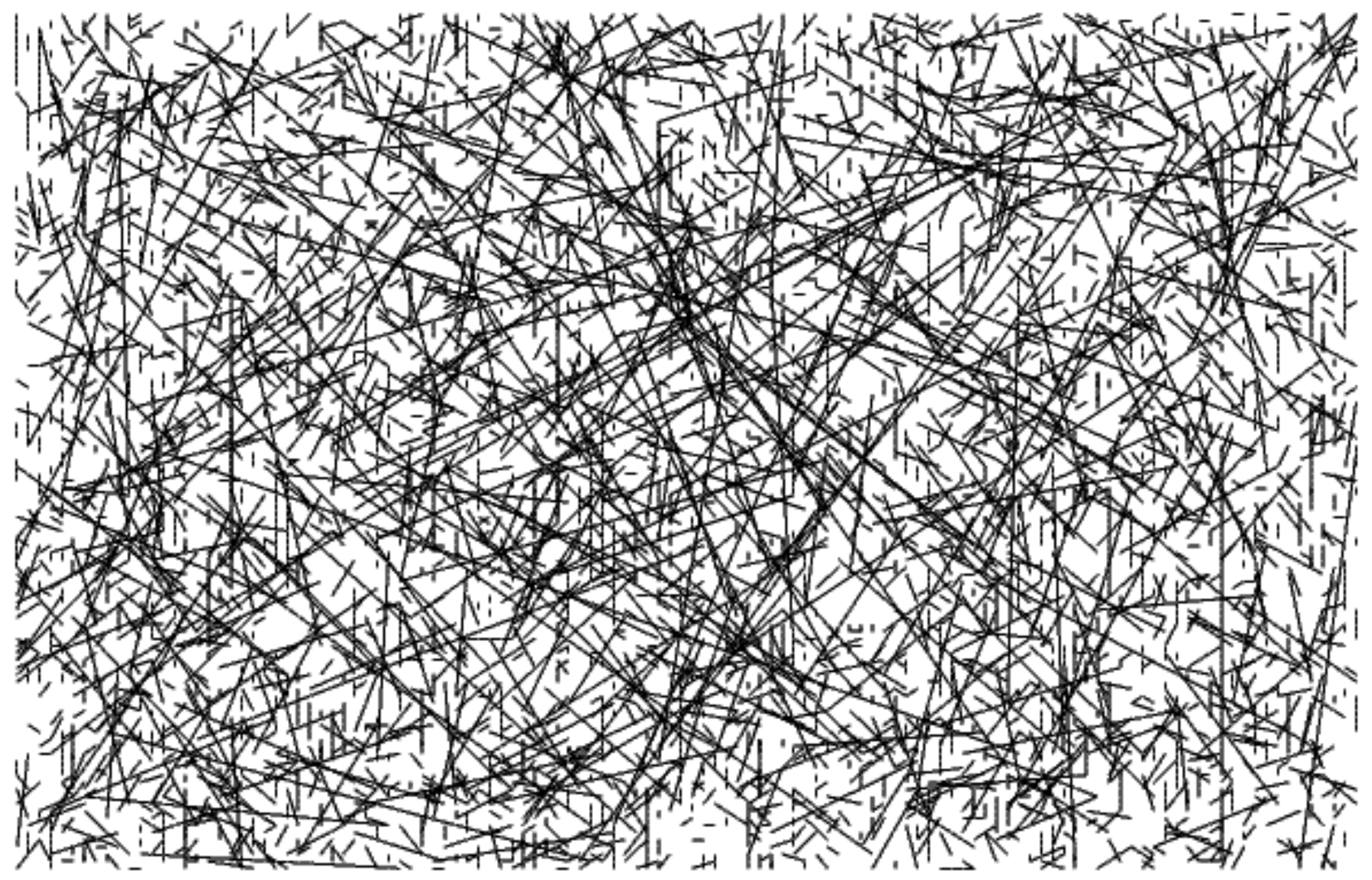}
}
\caption{\label{fig:red} Schematic representation of a city network. The city 
is composed by $120~\times~120~$blocks placed on a square grid. They are linked 
by (a) four short links connecting neighboring nodes and (b) long links 
connecting far away nodes (see text for more details). The city corresponds to 
the union of these two sub-networks.} 
\end{figure*}

Each node is linked with its first neighbors by a short link, as we illustrate 
in Fig.~\ref{fig:red_cortos}. They represent the human mobility pattern 
between all neighboring blocks. Notice that these links generate 
a connected graph. As we will see later, this characteristic of the network will 
allow the disease to reach all parts of the city (as long as it is not 
locked during the epidemic).   \\

On the other side, recent investigation on human mobility shows that the 
traveling lengths of the individuals follows a Levy distribution given by 
\cite{Gonzalez}

\begin{equation}
P(r)\propto (r+r_0)^{-\beta}
\end{equation}

\noindent where $P(r)$ stands for the probability that an individual reaches a 
distance $r$, and $r_0=100\,$m and $\beta=1.75$ correspond to empirical 
parameters. \\ 

We built a network of long links following the Levy mobility pattern. This 
network is illustrated in Fig.~\ref{fig:red_largos}. The procedure was as 
follows

\begin{enumerate}
 \item We randomly chose a node.
 \item We then select randomly (according to the Levy distribution) the length 
$r$ of the next link.
\item We link the node from step 1 to any (random) node located at the distance 
$r$ in any direction.
\end{enumerate}

Notice that each node may have more than one \textit{long link}. Recall that 
these links are complementary to the short links connecting neighboring blocks 
(say, four neighbors per block). \\

The total number of long links will depend on how many times we repeat the 
above steps. Thus, we looped these steps until the geodesic path of the network 
resembled the one expected for a ``small world'' network \cite{Barmak}. As a 
general rule, we found that this condition was fulfilled after $25\%$ of the 
nodes had at least one long link. Table~\ref{tab:coef_delays} summarizes the 
final set of parameter.\\

Finally, we stress the fact that the links between different blocks are assigned 
at the beginning of the simulation. Thereafter, the links remain fixed until the 
end of the simulation. \\

\begin{table}
\caption{Most relevant parameters used in the simulation of COVID-19 epidemic 
along the mobility network.}
\centering 
\begin{center}
\begin{tabular}{cc}
 \hline
 Number of nodes (blocks) & 120$\times$120\\
  Population per block & 300 \\
 Short links & between first neighbors \\ 
 Long links & over $25\%$ of the nodes \\
 Total number of links & 31825 \\
 $r_0$ (Levy distribution) & 100$\,$m \\
 $\beta$ (Levy distribution) & 1.75 \\
\hline
\end{tabular}\label{tab:coef_delays}
\end{center}
\end{table}

\subsection{\label{sec:mobility_pattern}Mobility pattern}

We follow a similar scheme as in Ref.~\cite{Barmak,Barmak2} for the 
individuals' mobility. This means that we divide the population of each block 
into two groups: \textit{stationary} and \textit{non-stationary} individuals. 
The former is composed by infected individuals while the latter is made up of 
susceptible, exposed and removed individuals. \\

\begin{figure}[!ht]
\centering
\includegraphics[width=0.7\columnwidth]{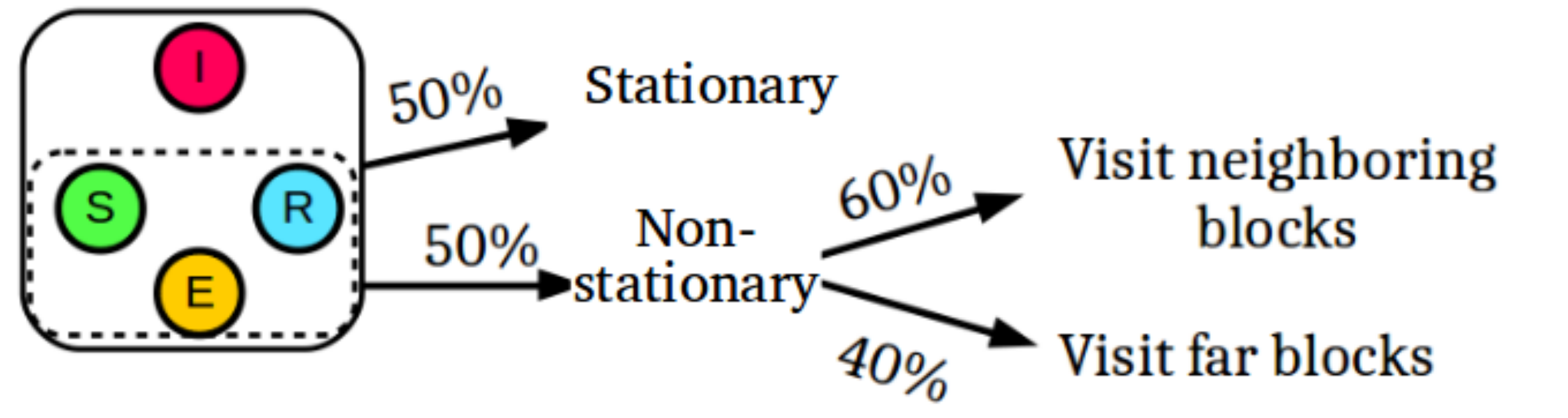}
\caption{\label{fig:active_passive} Schematic representation of 
the behavior of the population for a single block. The population is composed 
by susceptible (S), exposed (E), infected (I) and removed (R) individuals. 
Infected individuals stay at home during the simulation process while the 
$50\%$ of the susceptible, exposed and removed individuals move to another 
blocks (see text for more details).}
\end{figure}

We will assume that \textit{stationary} individuals stay in their original 
block. We will further assume, in the spirit of Ref.~\cite{Barmak2}, that $50\%$ 
of the \textit{non-stationary} stay in their original block, while the other 
$50\%$ is considered to be mobile during the simulation. In this sense, we 
assume that $60\%$ of the \textit{non-stationary} individuals travel (via short 
links) from their original block to the neighboring block. And, the remaining 
$40\%$ travel (via long links) from their original block to far blocks as 
indicated in Fig.~\ref{fig:active_passive}. \\

\subsection*{\label{sec:dinamics_individuals}Daily displacements of the 
individuals}

Each moving individual is assumed to stay 1/2 of the day in its original block 
while the other 1/2 of the day he (she) moves to another block. They go to his 
(her) destination (work) everyday, and at the end of the day return back to 
their original block (home). This mechanism is performed reversing the direction 
of the links between blocks. \\

\begin{figure*}[!ht]
\subfloat[\scriptsize{Home (0-12$\,$h)}\label{fig:inicio_final}]{
\includegraphics[scale=0.44]{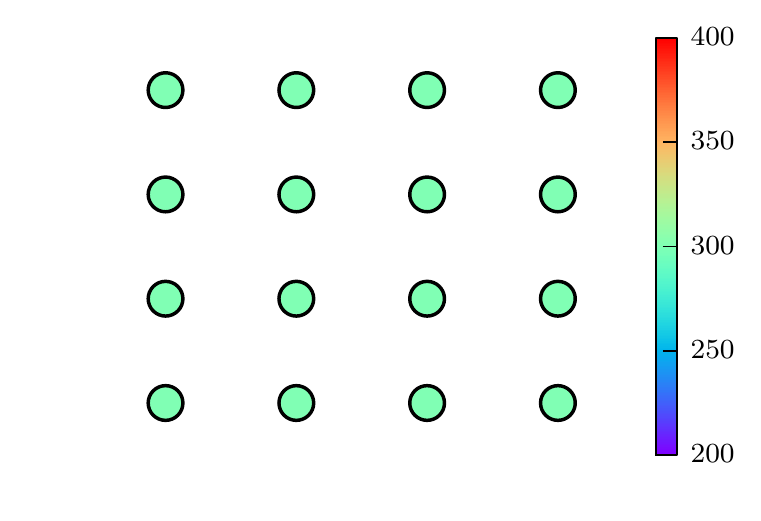}
}
\hspace{-5mm}
\subfloat[\scriptsize{Traveling to work}\label{fig:ida_flujo}]{
\includegraphics[scale=0.44]{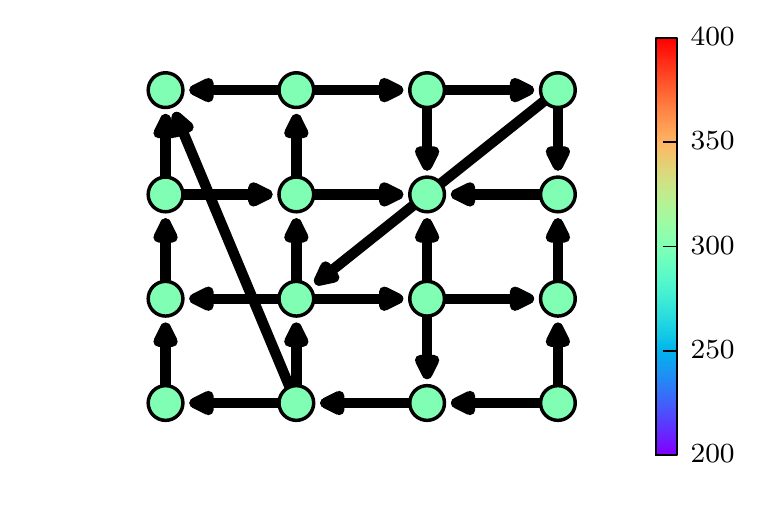}
}
\hspace{-5mm}
\subfloat[\scriptsize{Work (12-24$\,$h)}\label{fig:medio}]{
\includegraphics[scale=0.44]{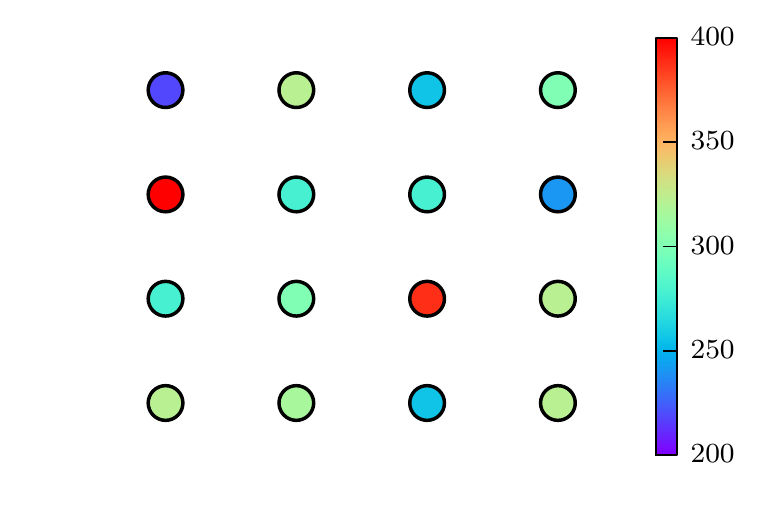}
}
\hspace{-5mm}
\subfloat[\scriptsize{Returning to home}\label{fig:regreso_flujo}]{
\includegraphics[scale=0.44]{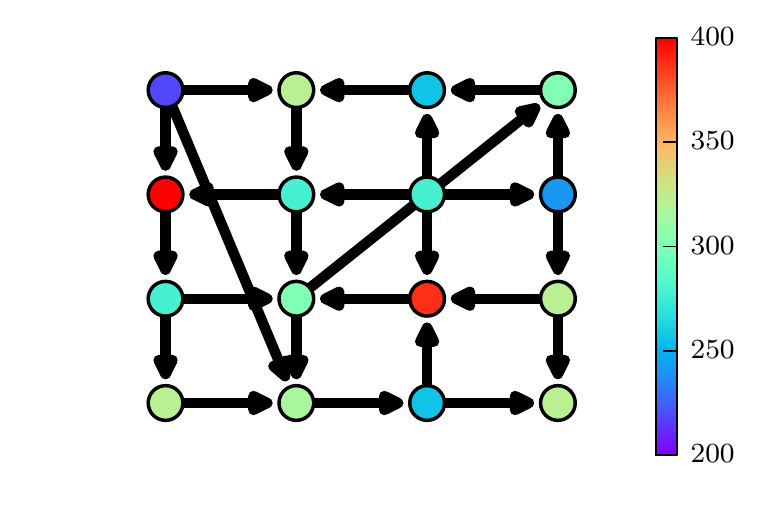}
}
\caption{\label{fig:evolucion} (Color on-line only) Schematic illustration of 
the epidemic model incorporating individual human mobility during each day. The 
node colors represent the population of each block (see scale bar on the 
right). (a) The day starts with all individuals at home. (b) At 12 o'clock, 
$50\%$ of the population of each block moves to other cells, according to the 
pre-assigned mobility pattern. The arrows represent the movement direction. (c) 
The population of each block after mobile humans traveled to the working block. 
(d) Then, at 24 o'clock those humans that traveled from home to work return 
back to home. At the end of the day, the population of each block is the same 
as that in the beginning of the day. The sequence (a-d) repeats every day until 
the end of the simulation.} 
\end{figure*}

Fig.~\ref{fig:evolucion} shows a schematic representation of the human mobility 
pattern during each day (we illustrate this by a small city of 
4$\times4~$blocks for a better visualization). Recall from the last section 
that the links represent the mobility pattern between different blocks. Thus, 
moving humans travel through the links from one block to another one. \\

\subsection{\label{sec:seir}Epidemiological dynamics of the 
individuals}

In order to describe the time evolution of a given population when a fraction 
gets infected, we resort to the SEIR compartmental models. These models 
consider that the individuals can be in four successive states: susceptible 
(S), exposed (E), infected (I) and removed (R). Details on each state can be 
found in Refs.~\cite{Anderson,Kermack,Barmak,Barmak2,cornes2020}. The 
``exposed'' state appears whenever the disease undergoes an 
``incubation'' period, as occurs in the context of the COVID-19. 
The ``removed'' state includes either recovered and dead people. \\

The equations that describe the evolution of the infection read as follows 
\cite{Anderson,Kermack}

\begin{equation}
\left\{\begin{array}{ccc}
\dot{s}(t) & = & -\beta i(t)s(t)\\
\dot{e}(t) & = & \beta i(t)s(t)-\sigma e(t)\\
\dot{i}(t) & = & \sigma e(t)-\gamma i(t)\\
\dot{r}(t) & = & \gamma i(t)\\
s(t)+e(t)+i(t)+r(t) & = & 1 \\
\end{array}\right.\label{eq:seir}
\end{equation}\\
    
\noindent where $s(t)=S(t)/N$, $e(t)=E(t)/N$, etc. correspond to the fraction of 
people in  each state. For the purpose of simplicity, we will consider the 
coefficients $\beta$, $\sigma$ y $\gamma$ as fixed parameters. The parameter 
$\beta$ (infection rate) depends on intrinsic ingredients like the infectivity 
of the virus under consideration and extrinsic ones like the contact 
frequency. Besides, the parameters $\sigma$ and $\gamma$ depend exclusively on 
the illness under consideration. \\

The basic reproduction number $R_0$ is defined at the beginning of the 
propagation process as \cite{Dushoff}
 
\begin{equation}
    R_0 = \frac{\beta}{\gamma}\label{eq:r0}
    \end{equation}
    
This quantity represents the number of individuals that are infected by the 
first infected individuals (say, at the beginning of the disease). It is 
straight forward that the infection will blossom if $R_0$ is larger than 1. On 
the contrary, if this quantity is smaller than 1 the infection vanishes. \\

\begin{figure*}[!ht]
\centering
\subfloat[\scriptsize{basic SEIR model}\label{fig:seir_basico}]
{\includegraphics[scale=0.8]{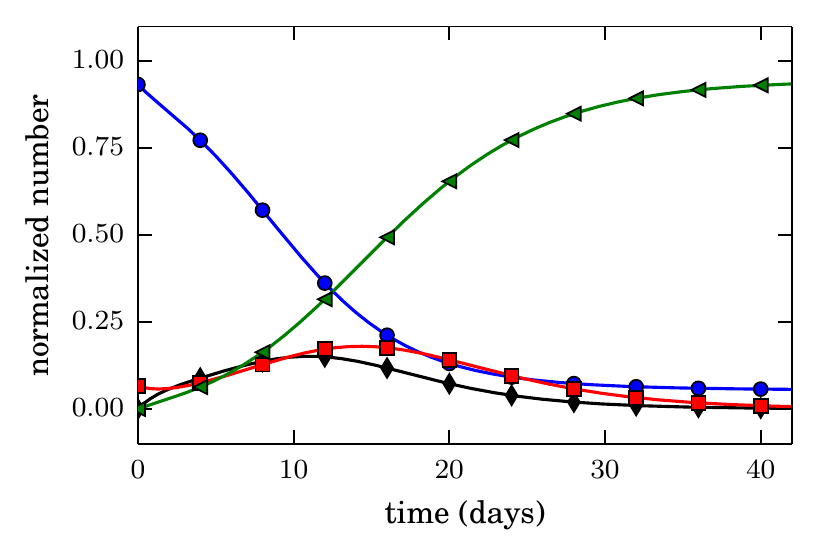}
}
\subfloat[\scriptsize{spatio-temporal SEIR model}\label{fig:seir_espacial}
] {\includegraphics[scale=0.8]{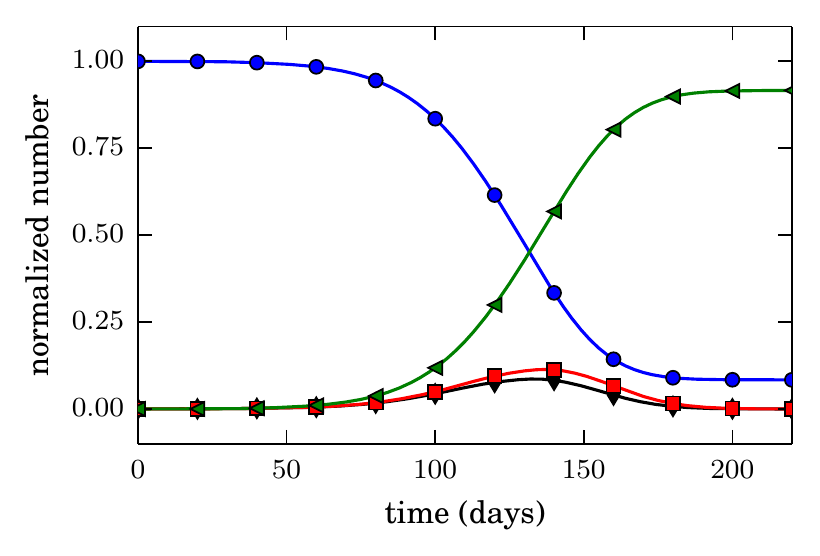}
}
\caption{\label{fig:seir_comparacion} Normalized number of susceptibles
(\protect\tikz\protect\draw[black,fill=blue] (0,0) circle (.9ex);), exposed
(\mydiamond{black}), infected (\protect\tikz\protect\draw[black,fill=red]
(0,0) rectangle (0.2cm,0.2cm);) and removed (\mytriangleup{green}) as a
function of time. In (a) we simulated a single block, while in (b) the 
scenario corresponds to a city composed by $120~\times~120~$blocks placed on a 
square grid (see Section \ref{sec:scenario}). The plots are normalized with 
respect to the total population in the city: (a) 300 and (b) $4.32\,$M 
(\textit{i.e.} 300 individuals per block). In both cases, the disease spreads 
without any kind of intervention strategy. The infection rate equals 
$\beta_0=0.75$ all along the propagation process. The simulation started with 
20 infected individuals located in the single block (a) and at the central 
block of the city (b). The rest of the individuals were assumed to be in the 
susceptible state.}
\end{figure*}

Fig.~\ref{fig:seir_comparacion} shows the time evolution of 
either the SEIR model for a single block and for the whole mobility model 
(see caption for details). Notice that the latter shifts the date for the 
maximum infection to approximately 150. We will analyze this behavior in more 
detail in Section \ref{sec:results}. \\

\subsection{\label{sec:lockdown_types}Lockdown types}

In this section we define the proposed containment strategies which we have 
found useful in order to mitigate the disease. We divide the different types of 
quarantine according to their level of real life implementation difficulty

\begin{itemize}
 \item Global lockdown
 \item Imperfect lockdown
 \item Local lockdown
\end{itemize}

\subsection*{\label{sec:global_lockdown}Global lockdown scenario (GLs)}

This type of lockdown consists in the isolation of each block. This means 
that people remain in their ``home'' blocks. We stress the fact that, within 
this strategy, society as a whole enters in lockdown. We illustrate this 
containment action in Fig.~\ref{fig:diagrama_cuarentena_global}.\\

\subsection*{\label{sec:imperfect_lockdown}Imperfect lockdown scenario (IGLs)}

As in the previous case, the whole society adopts the same behavior. But, in 
this mitigation strategy, we partially reduce the movement between different 
blocks. Recall from Section~\ref{sec:mobility_pattern} that $50\%$ of the 
\textit{non-stationary} individuals move from their original block to another 
one. In this strategy, we reduce the level of mobility between blocks. Notice 
that the global quarantine corresponds to a full reduction of the mobility. We 
scheme this containment action in 
Fig.~\ref{fig:diagrama_cuarentena_imperfecta}.\\

\subsection*{\label{sec:local_lockdown}Local lockdown scenario (LLs)}

\begin{figure*}[!ht]
\subfloat[\scriptsize{Without lockdown}\label{fig:diagrama_sin_cuarentena}]{
\includegraphics[scale=0.14]{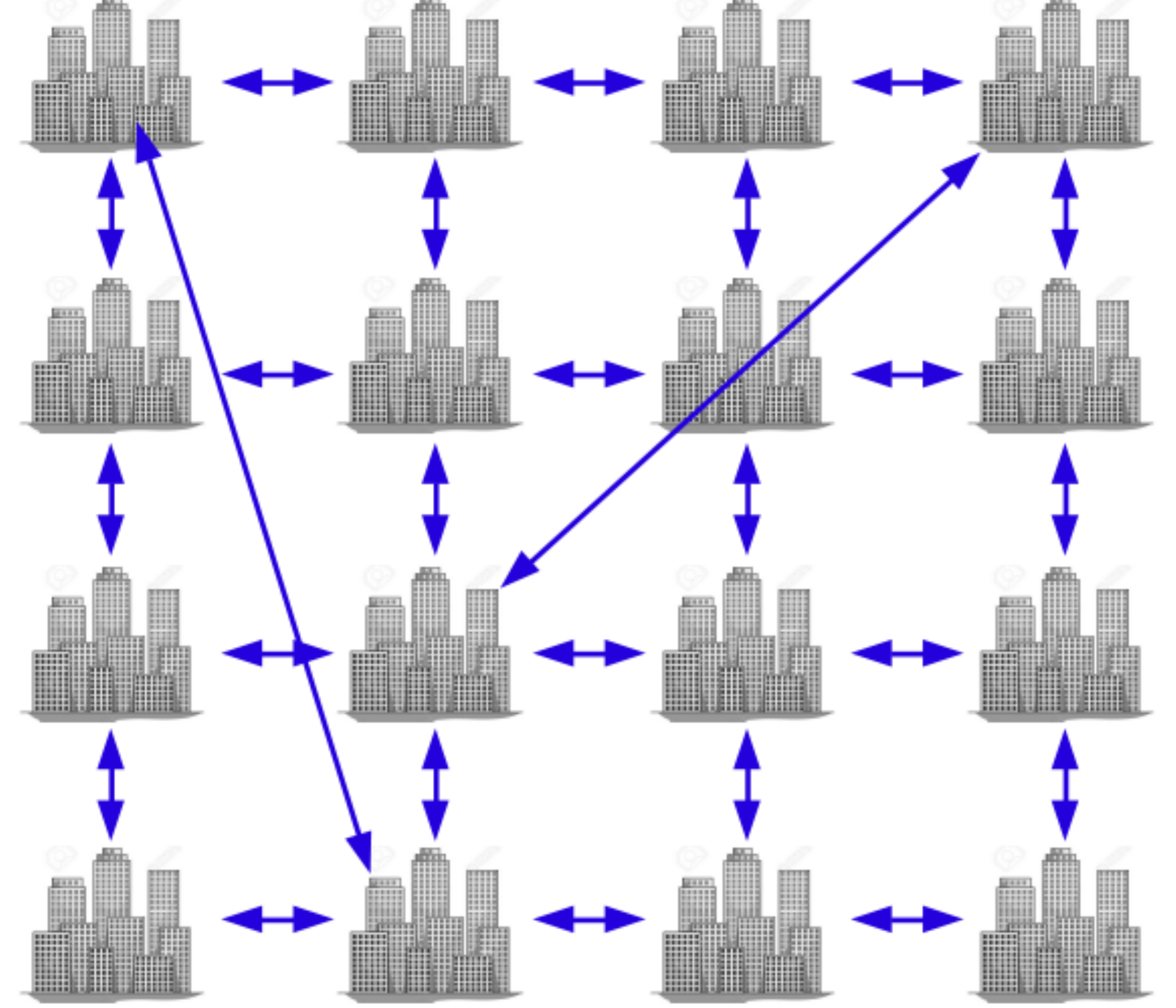}
}
\hspace{2mm}
\subfloat[\scriptsize{Global lockdown}\label{fig:diagrama_cuarentena_global}]{
\includegraphics[scale=0.14]{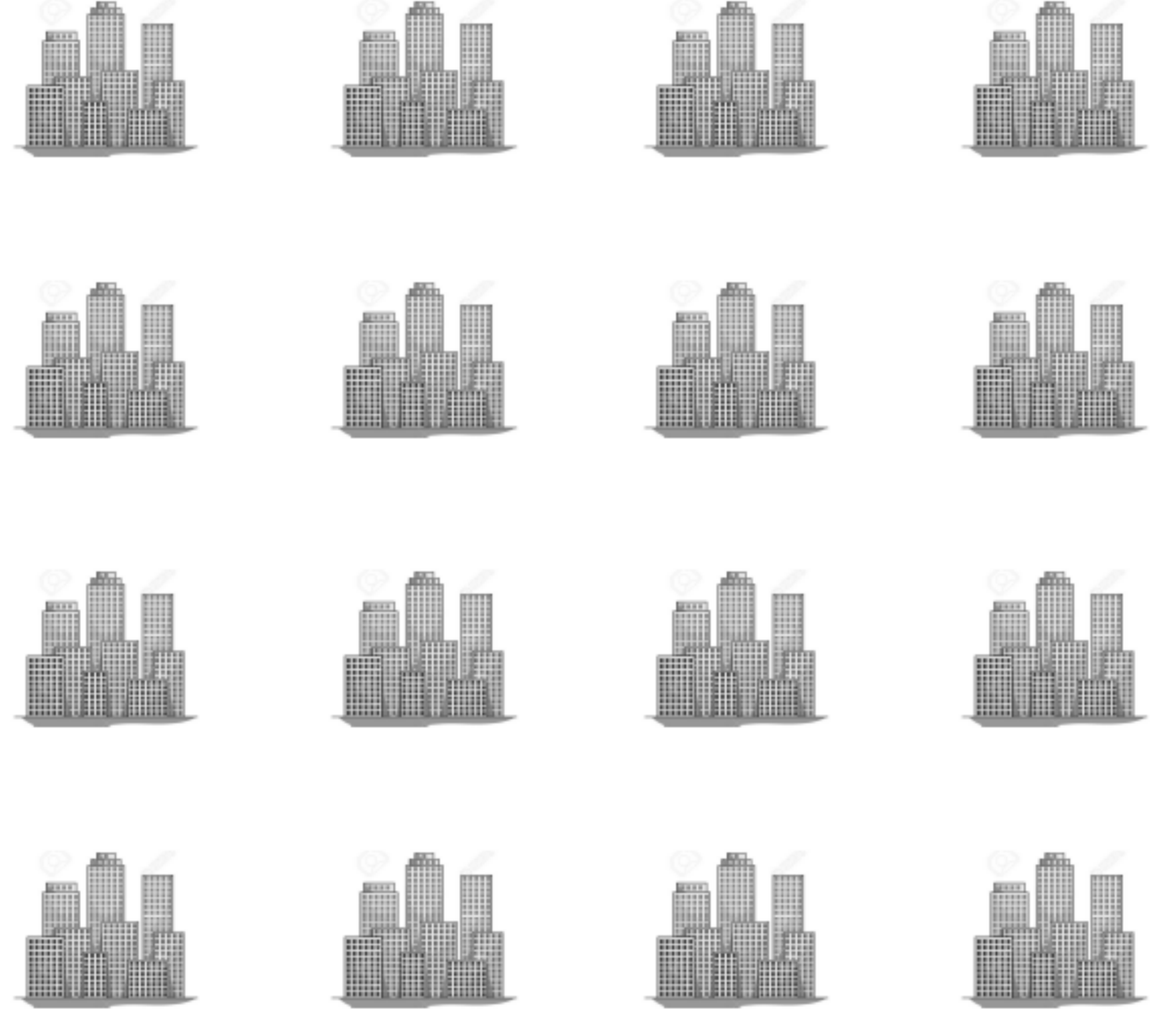}
}
\hspace{2mm}
\subfloat[\scriptsize{Imperfect 
lockdown}\label{fig:diagrama_cuarentena_imperfecta}]{
\includegraphics[scale=0.14]{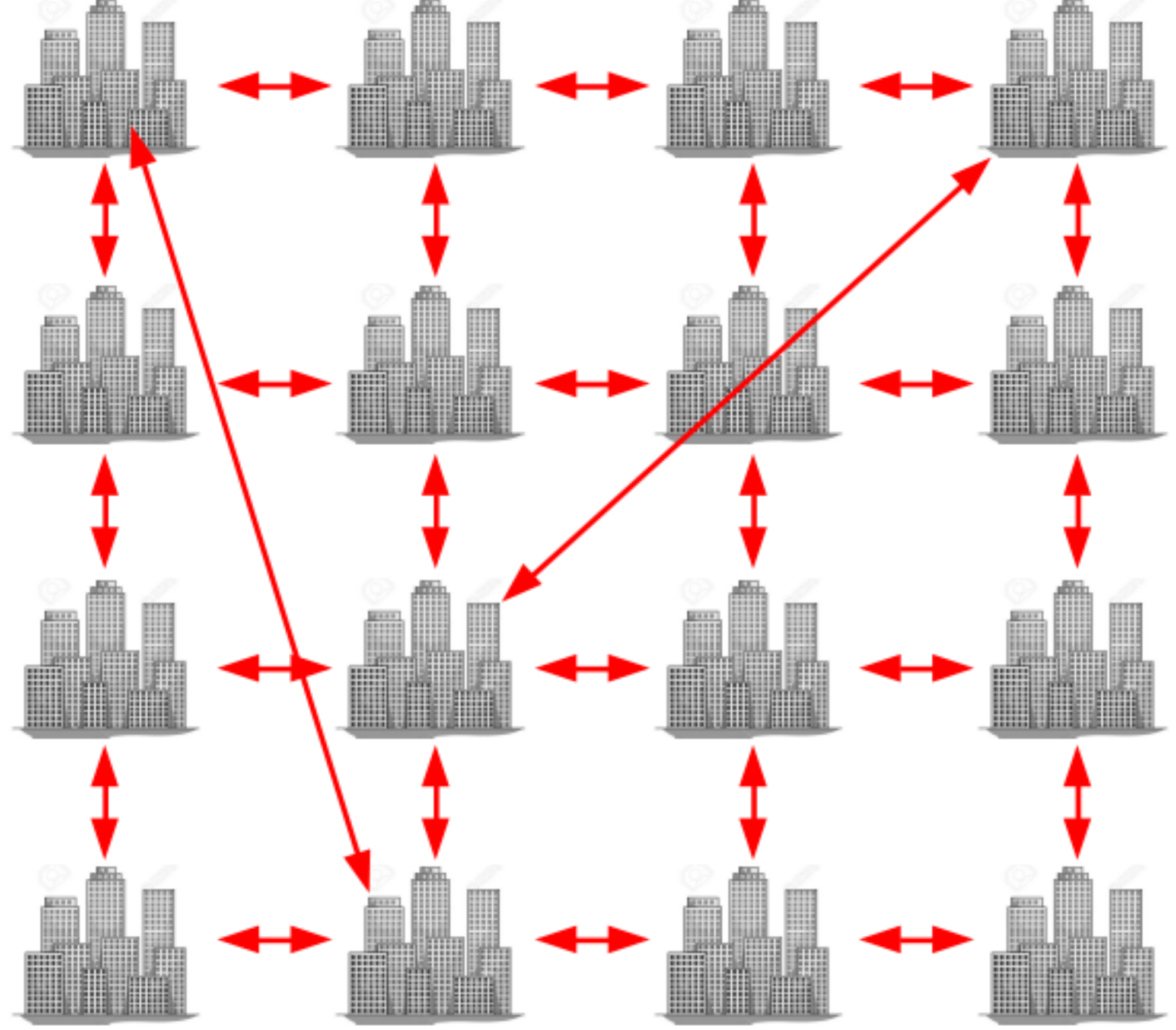}
}
\hspace{2mm}
\subfloat[\scriptsize{Local lockdown}\label{fig:diagrama_cuarentena_local}]{
\includegraphics[scale=0.14]{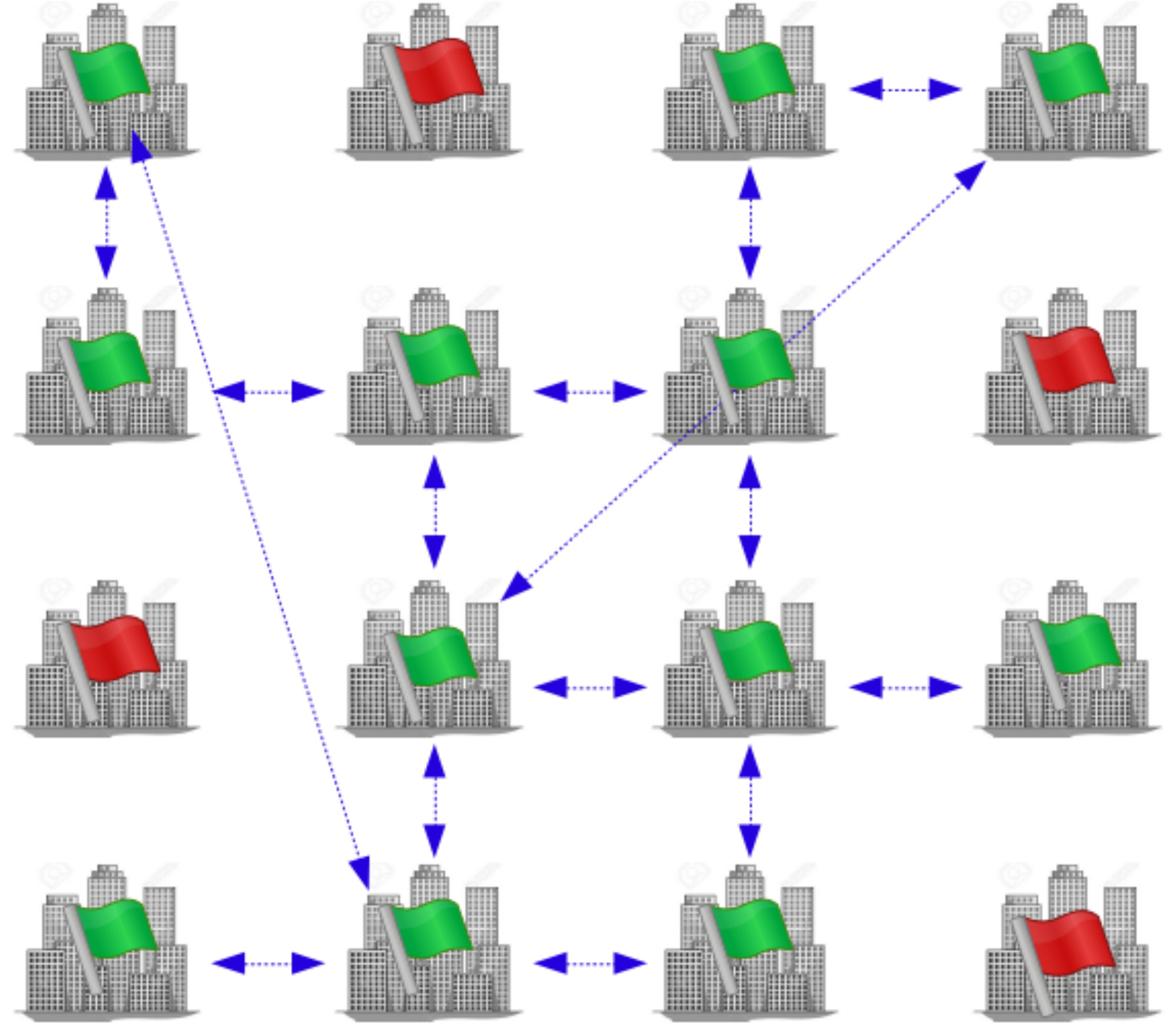}
}
\caption{\label{fig:tipos_cuarentena} (a-d) Schematic representation of 
different types of mitigation strategies (exemplified by  a $4\times4$ grid). 
The lines represent the human mobility pattern between blocks. (a) People are 
allowed to evolve freely, equivalent to a ``continuous activity'' scenario. That 
is, there is no intervention during the epidemic. (b-d) Before a given day, 
individuals move freely from one block to another according to the mobility 
pattern. After that, (b) all blocks are isolated, (c) the flow between blocks 
(represented by red arrows) is reduced but the mobility pattern is unaffected, 
(d) those blocks with a number of infected individuals greater than a certain 
threshold are isolated (labeled with a red flag). The other blocks remain 
linked to each other (labeled with a green flag). See text for more details.} 
\end{figure*}

Unlike the other lockdown types, this applies to certain blocks and not to the 
whole society. Only those blocks that have a certain number of infected people 
are isolated. In this sense, the population of the isolated block are 
prohibited to travel around the city. This is achieved in practice by 
``cutting'' the links to/from the infected blocks. We illustrate this 
containment action in 
Fig.~\ref{fig:diagrama_cuarentena_local}.\\

\section{\label{sec:simulations}Numerical simulations}

We integrate the SEIR equations by means of the Runge Kutta 4th-order method. 
The chosen time step was 0.1 (days). The SEIR equations 
were updated twice a day (after the people left their homes and after they 
returned back (see Figs.~\ref{fig:inicio_final} and \ref{fig:medio}).\\

As mentioned in Section~\ref{sec:seir}, the parameters $\sigma$ and 
$\gamma$ represent the incubation rate and the recuperation rate, 
respectively. Therefore, $\sigma^{-1}$ and $\gamma^{-1}$ correspond to the 
mean incubation time and the mean recovery time, respectively. According to 
preliminary estimations for COVID-19, we consider the following parameter 
values for the SEIR model: $\sigma^{-1}=3\,$days and 
$\gamma^{-1}=4\,$days \cite{Tian,Milo,oms,hopkins,Lessler}.\\

Infected individuals remain at their ``Home'' until evolving into the 
removed state. Susceptible, exposed and removed individuals are able to 
move from one block to another. The simulation started with 20 infected 
individuals located at the central block of the city, while the rest of the 
individuals were assumed to be in the susceptible state. \\

According to preliminary estimations for COVID-19, the basic reproduction number 
$R_0$ is (approximately) 3 \cite{Tian,ECDC,Liu}. This means that the infection 
rate $\beta_0$ is 0.75 (considering $\gamma^{-1}=4\,$days). The implementation 
of complementary health policies (use of mask, social distancing policies, among 
others) tends to reduce the contact frequency, and, therefore, the infection 
rate ($\beta$). Thus, we will also examine situations accomplishing infection 
rates of a fraction of $\beta_0$. \\

\section{\label{sec:results}Results}

We will examine three major scenarios affecting the human mobility:

\begin{enumerate}
 \item The (global) lockdown scenario assumes that people remain confined at 
home until the epidemic is almost over. See details in 
Section~\ref{sec:results_global_lockdown}. \\

 \item The scenario where the confinement recommendation is followed by a 
fraction of the city inhabitants. We assume that the traveling 
individuals move around according to the Levy pattern explained in 
Section~\ref{sec:epidemiological_model}. See further details on this scenario 
in Section~\ref{sec:results_imperfect_lockdown}. \\

 \item Mobility is suppressed only for the inhabitants of ``infected blocks''. 
That is, common life mobility is sustained between blocks where no symptoms of 
the disease appeared. See Section~\ref{sec:results_local_lockdown} for 
details. \\
\end{enumerate}

This last scenario is the most cumbersome one since ``non-symptomatic'' does 
not actually mean ``non-infected''. We will explicitly introduce a set of 
``non-symptomatic'' individuals in this scenario, in order to understand 
possible flaws to confinement. \\

\subsection{\label{sec:results_global_lockdown}Global lockdown scenario (GLs)}

The GLs means that people remain confined within the block where 
she (he) lives. It corresponds to a sudden break of the mobility 
around the city in the context of our model. We assume, however, that people 
may still get in contact within their own block. \\

In this case, we consider the mobility suppression as the only heath-care 
policy. Additional health-care recommendations for the every day living (say, 
masks, common rooms disinfections, etc.) are considered in 
\ref{sec:global_complementarias}. We will come back to this issue at the end of 
this Section. \\

\begin{figure*}[!ht]
\centering
\subfloat[\scriptsize{Infected}\label{fig:plot_rl_3_0_infectados}]{
\includegraphics[scale=0.8]{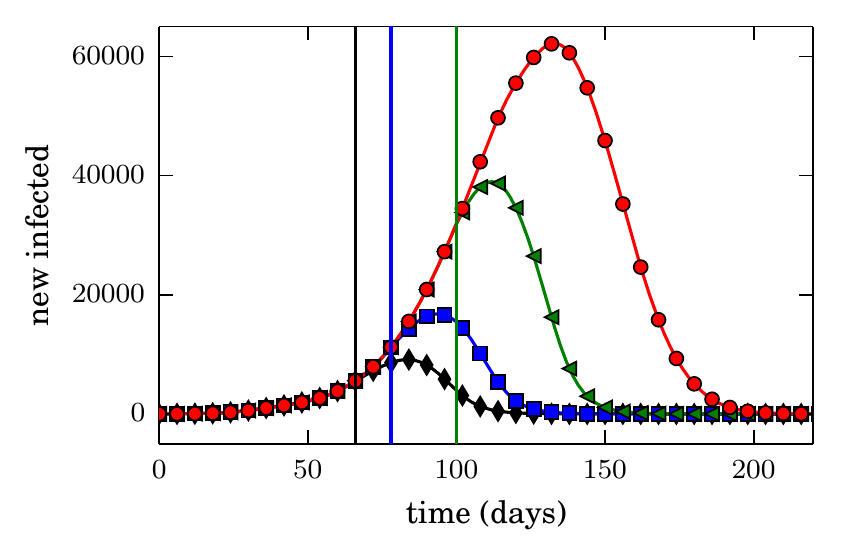}
}
\subfloat[\scriptsize{Removed}\label{fig:plot_rl_3_0_recuperados}]{
\includegraphics[scale=0.8]{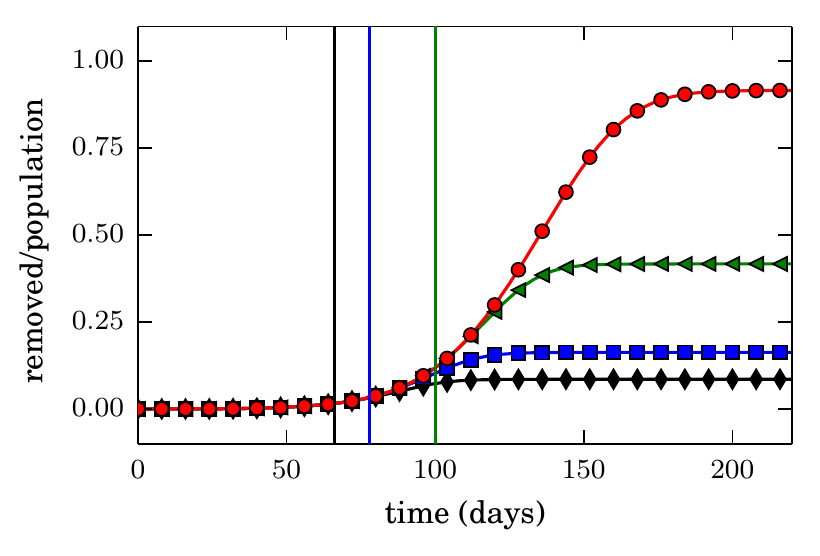}
}
\caption{\label{fig:infect_recup_global} (a) Number of new infected and (b) 
normalize number of removed as a function of time. The plot in (b) is 
normalized with respect to the total population in the city ($4.32\,$M). The 
GLs is applied when the number of new infected equals to: 
\mydiamond{black}~5k, \protect\tikz\protect\draw[black,fill=blue] (0,0) 
rectangle (0.3cm,0.3cm); 10k and \mytriangleleft{green}~30k new infected 
individuals. \protect\tikz\protect\draw[black,fill=red] (0,0) 
circle (.9ex); corresponds to the scenario of no lockdown at all. The 
infection rate remains constant along the simulation process and equals to 
$\beta_0=0.75$. That is, there is no complementary health-care policies during 
the lockdown. The different lockdown implementation days are indicated by 
vertical lines.} 
\end{figure*}

Fig.~\ref{fig:plot_rl_3_0_infectados} shows the number of new infected people 
along time for three different lockdown periods (see caption for details). It 
is also shown the evolution for the case of no lockdown at all. The mobility 
cutoff prevents the infection curves in Fig.~\ref{fig:plot_rl_3_0_infectados} 
from growing almost immediately after the beginning of the lockdown. The 
disease, however, disappears (approximately) 50 days (or 7 weeks) after. 
This is the time it take the susceptible or exposed individuals in each block to 
surpass the disease.\\

Fig.~\ref{fig:plot_rl_3_0_recuperados} exhibits the number of 
removed individuals as a function of time (see caption for details). These 
correspond to those individuals that previously appear as infected in 
Fig.~\ref{fig:plot_rl_3_0_infectados}. It can be seen that the number of 
removed individuals increases since the outbreak of the disease, but it 
reaches a plateau soon after the lockdown is established. The plateau level, 
however, depends strongly  on the starting date of the lockdown. 
Recall that the disease evolves only within the infected blocks after the 
lockdown implementation. Thus, the earlier the lockdown, the less number of 
infected (and removed) blocks at the end of the disease.  \\

\begin{center}
\begin{figure}
\subfloat[Infected\label{fig:infec_global_snap}]{
\includegraphics[scale=0.8]{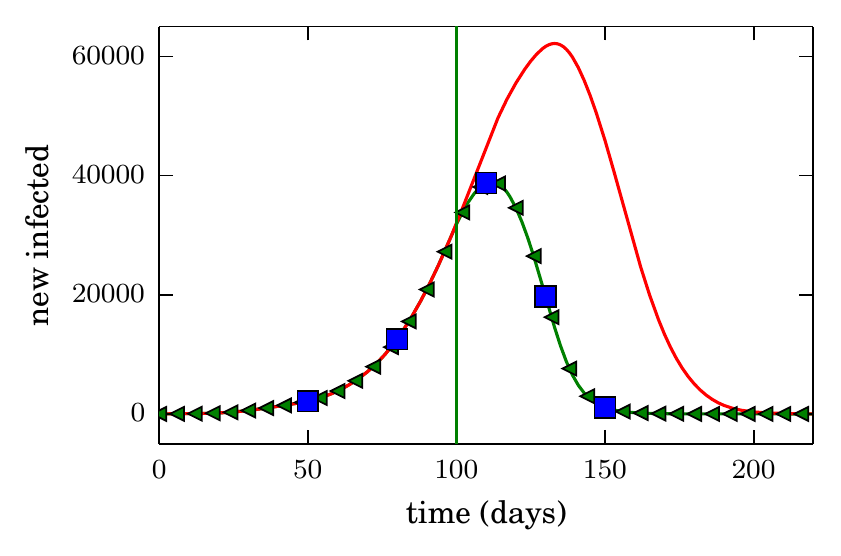}
}
\subfloat[Removed\label{fig:recup_global_snap}]{
\includegraphics[scale=0.8]{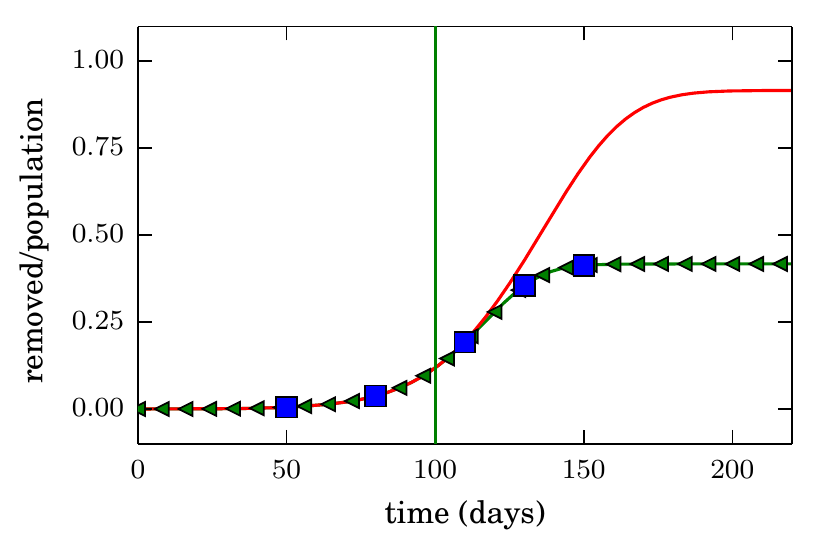}
}\\
\begin{tabular}{@{\hspace{0mm}}c|@{\hspace{0mm}}c|@{\hspace{0mm}}c|@{\hspace{0mm
}}c|@{\hspace{0mm}}c}
\hline
\multicolumn{5}{c}{Infected}\\
\hline
\includegraphics[scale = 0.11]{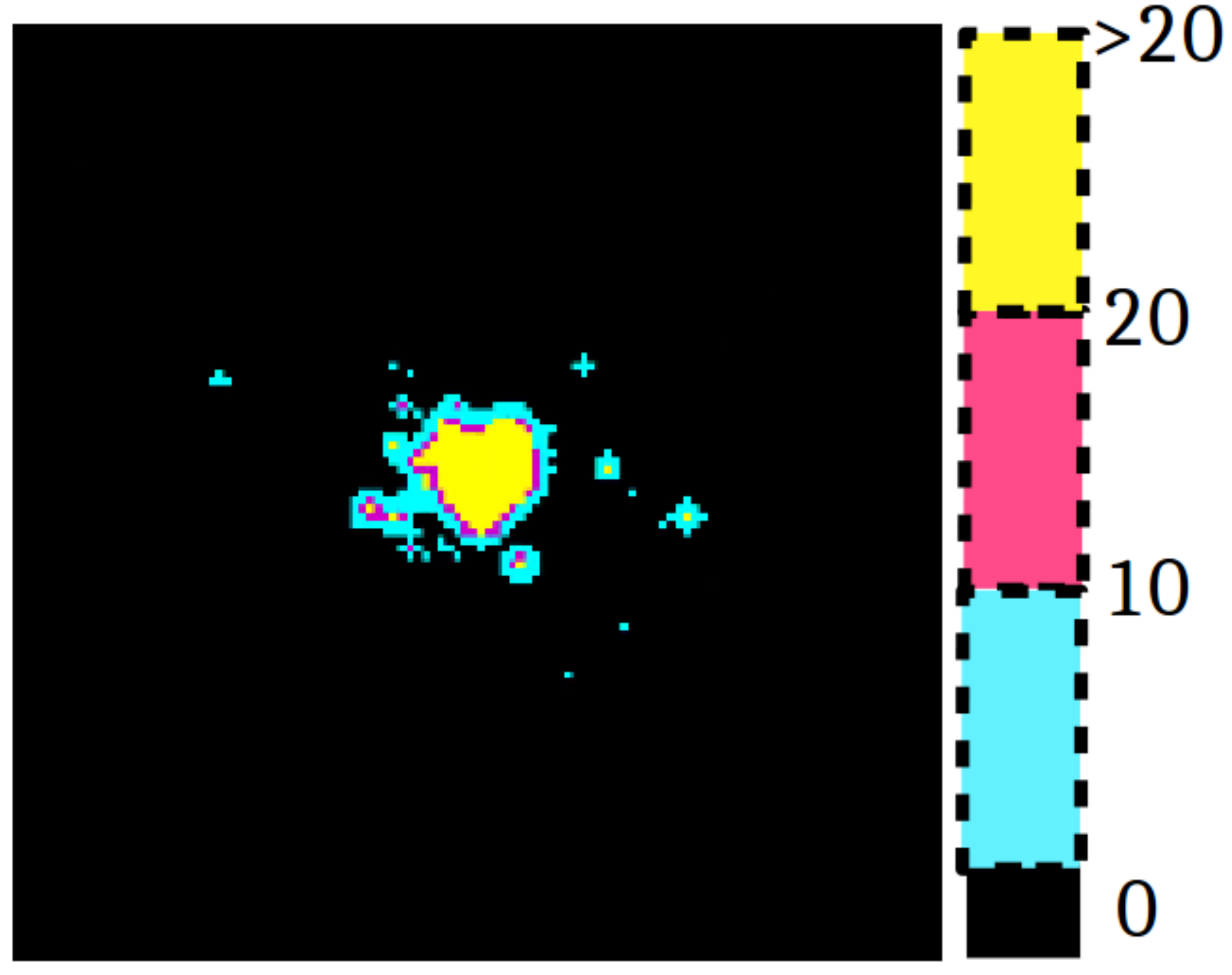} & 
\includegraphics[scale = 0.11]{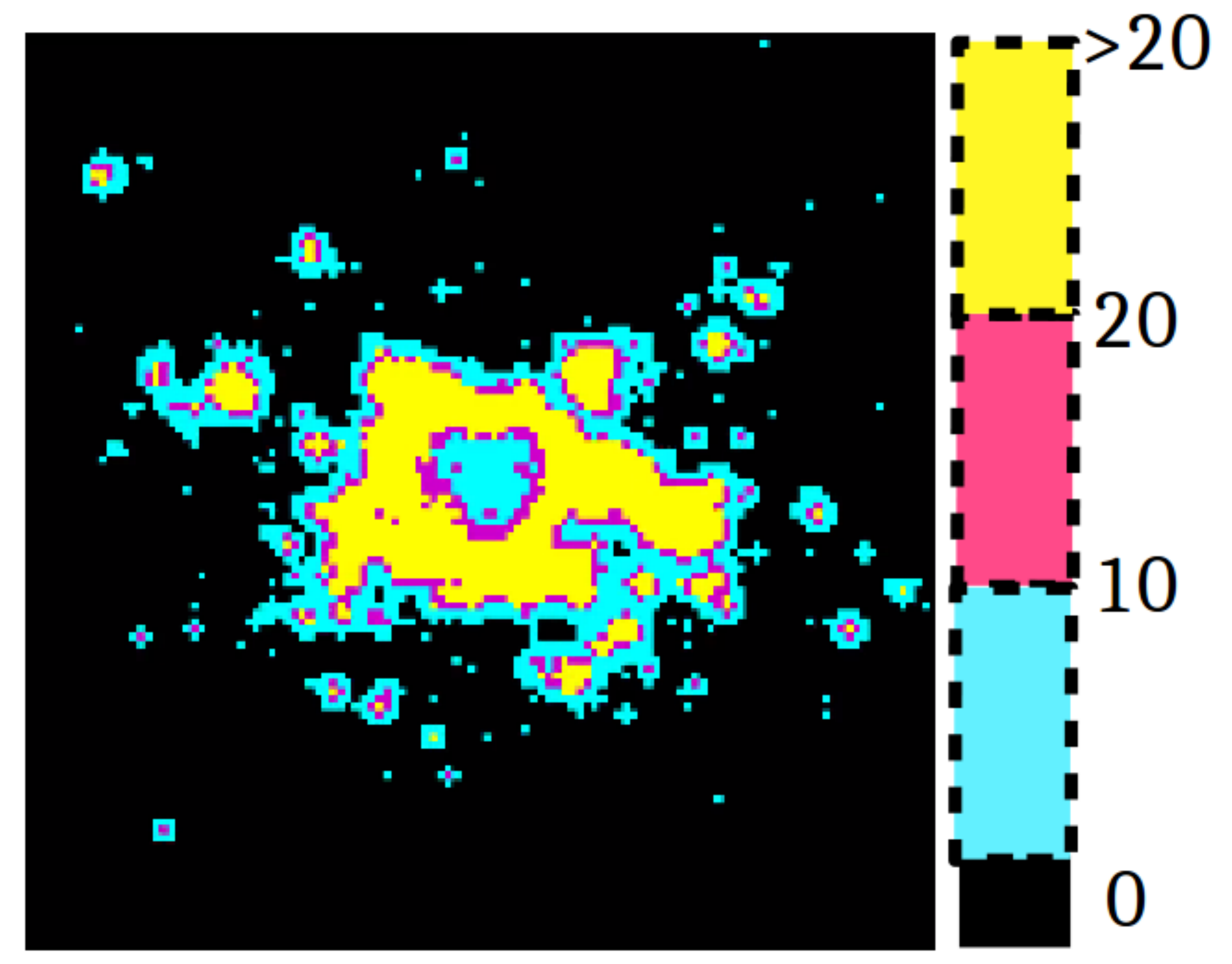} & 
\includegraphics[scale = 0.11]{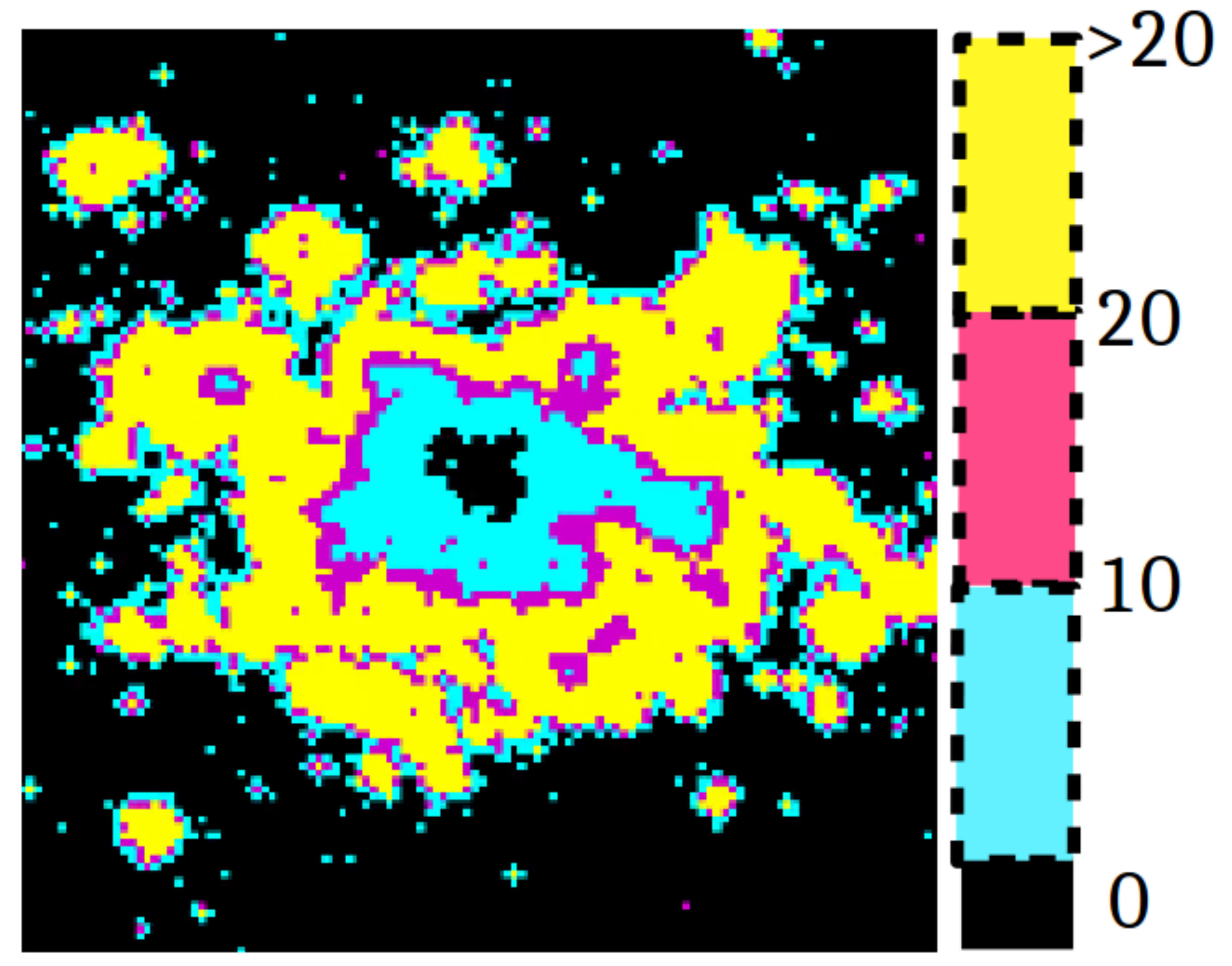} & 
\includegraphics[scale = 0.11]{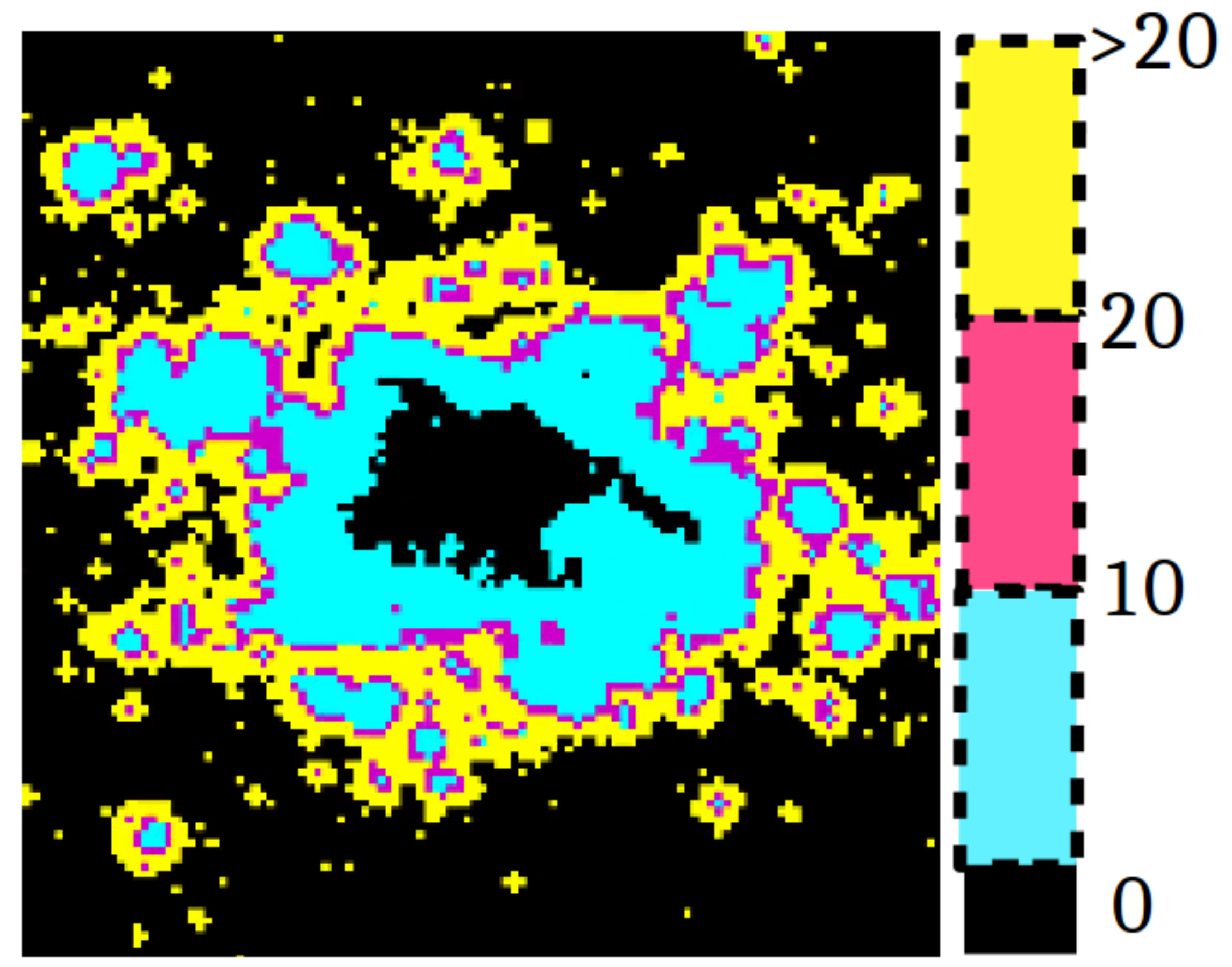} & 
\includegraphics[scale = 0.11]{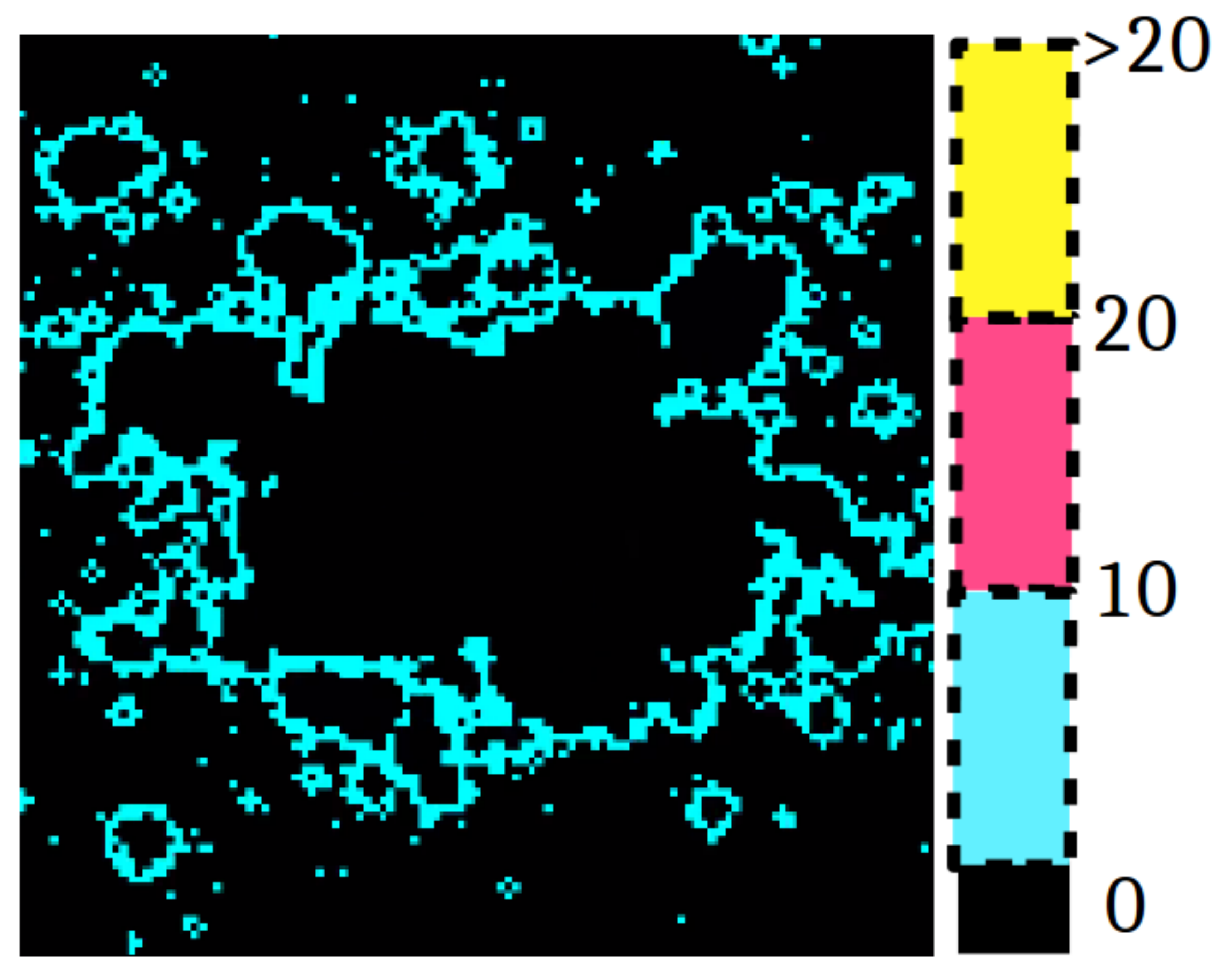} 
  \\\hline
  \multicolumn{5}{c}{Removed}\\
  \hline
\includegraphics[scale = 0.11]{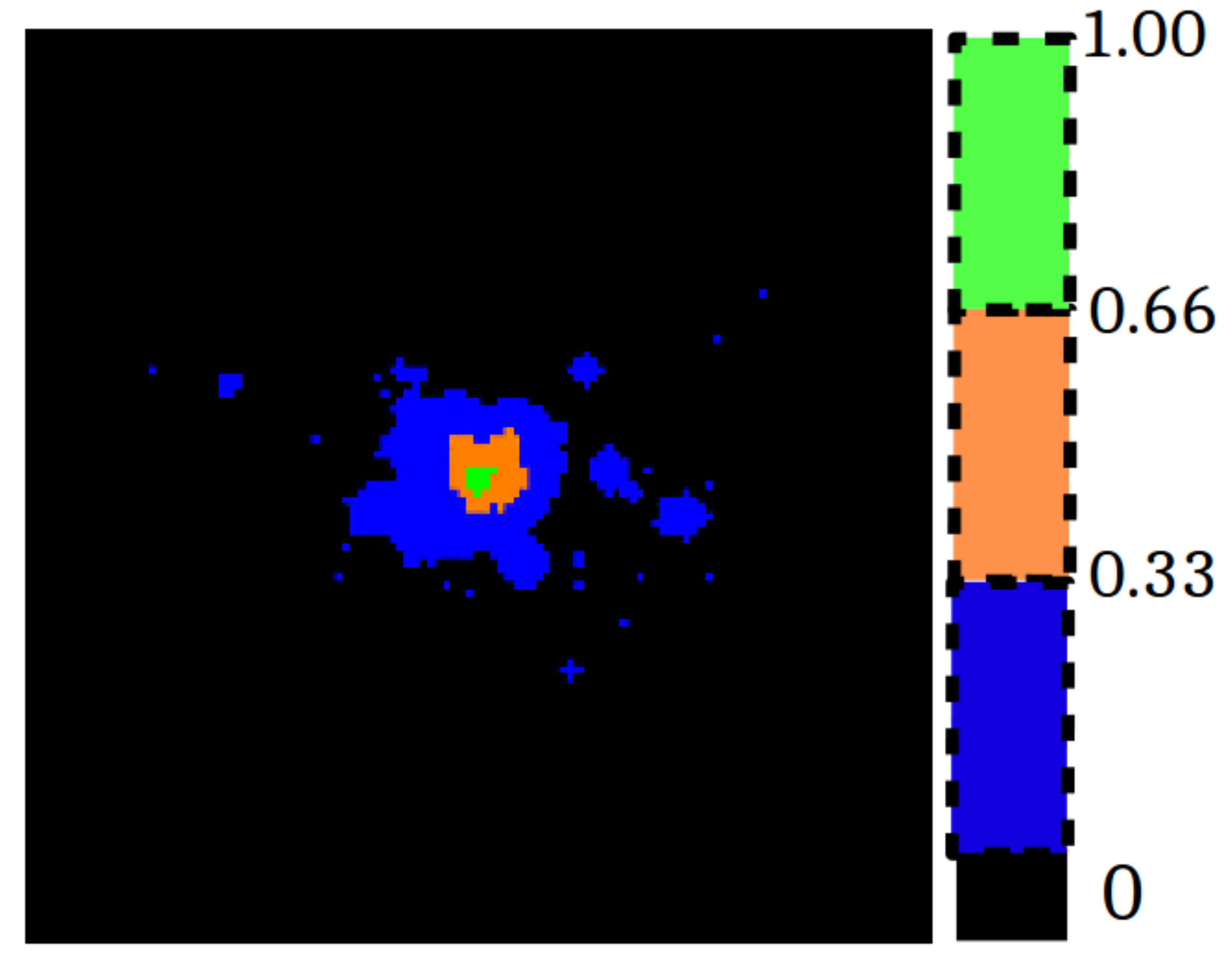} & 
\includegraphics[scale = 0.11]{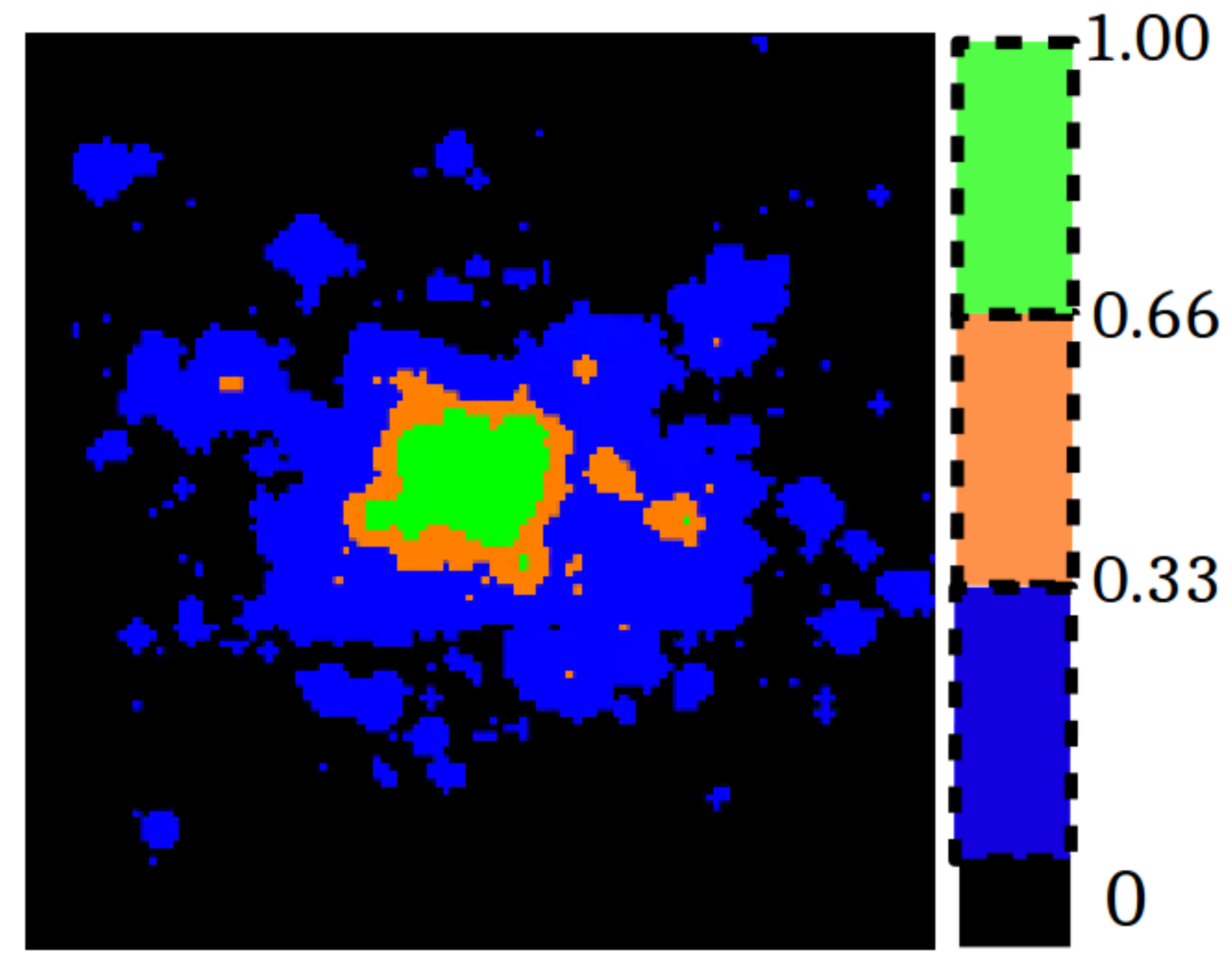} & 
\includegraphics[scale = 0.11]{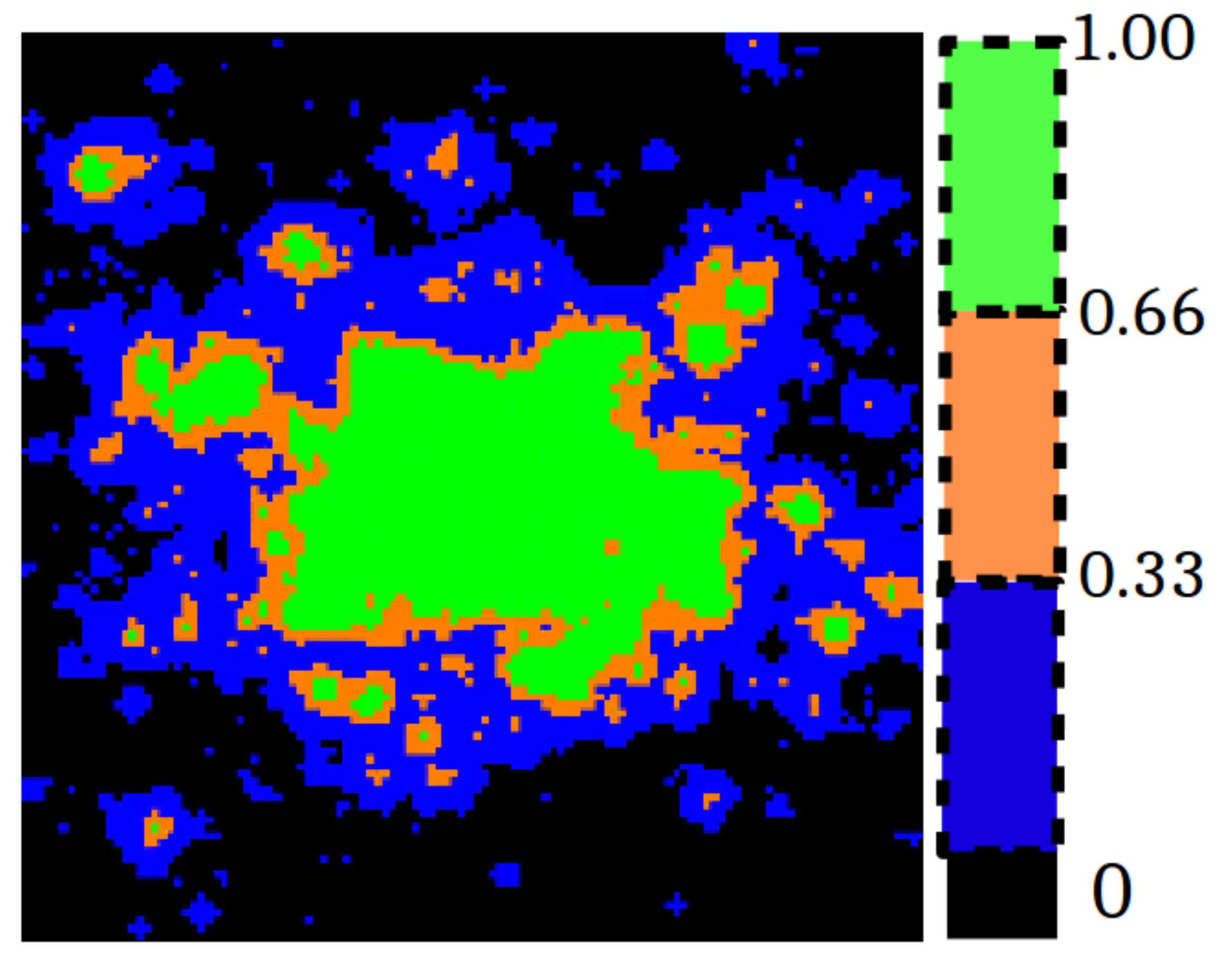} & 
\includegraphics[scale = 0.11]{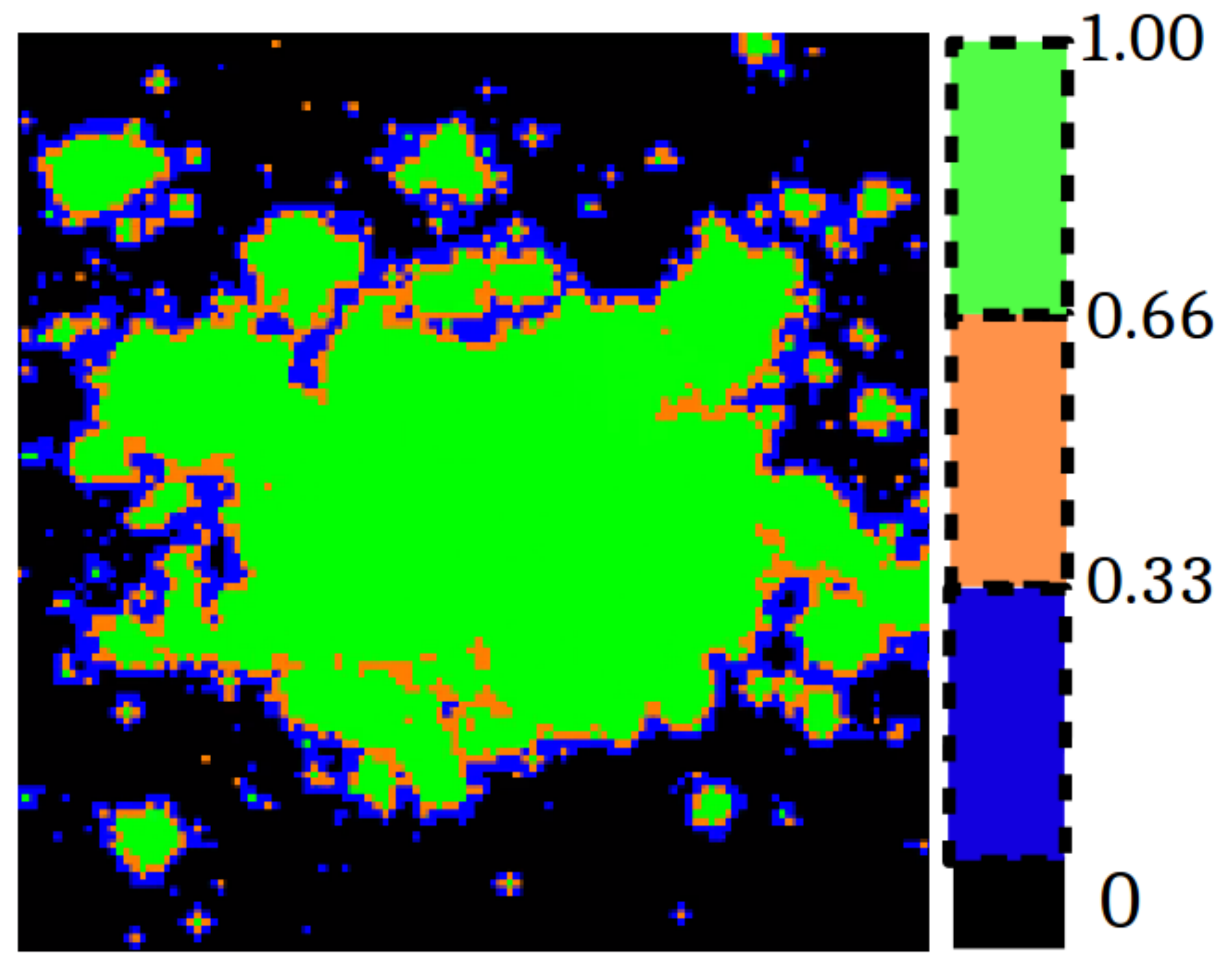} & 
\includegraphics[scale = 0.11]{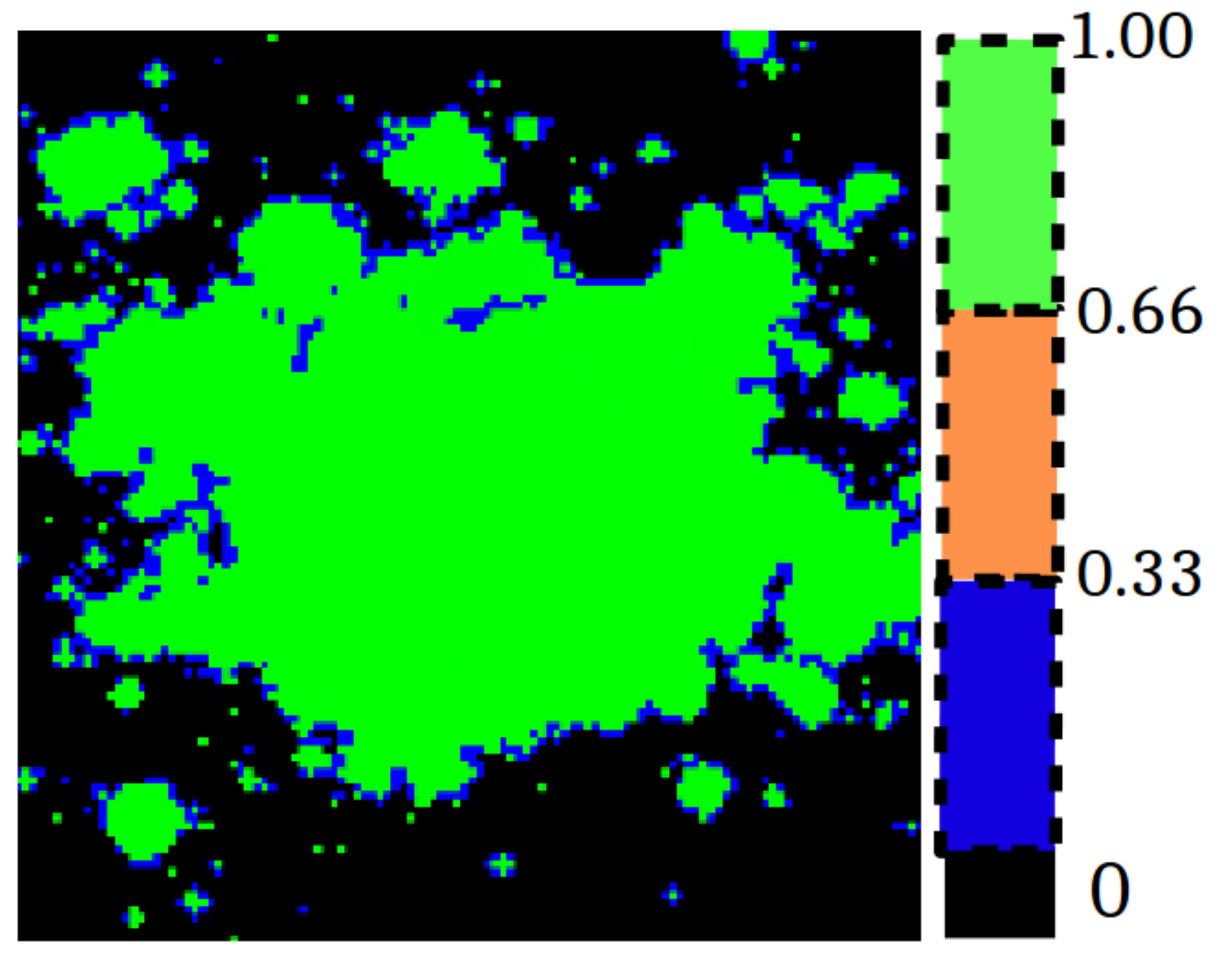}\\
\hline
Day 50 & Day 80 & Day 110 & Day 130 & Day 150 \\
\hline
\end{tabular}
\caption{(a) Total number of new infected and (b) normalized number of 
removed as a function of time. The plot in (b) is normalized with respect to 
the total population in the city ($4.32\,$M). The GLs started at the 
day 100 after the first infected was detected. This is indicated by a vertical 
line in (a) and (b). The red continuous line corresponds to the scenario of no 
lockdown at all, while the lockdown case is indicated by green triangles. 
(Lower) Spatial distribution of the number of infected individuals for the five 
different dates (indicated in (a) and (b) by blue squares). The scale bar on 
the right corresponds to the number of infected and the normalized number of 
removed per block, respectively. The normalization was computed with respect 
to the population of each block. The infection rate remains constant along 
the simulation process and equals $\beta_0=0.75$. The city was composed by 
$120\times120~$blocks placed on a square grid with 300 individuals per block.}
\label{tab:snapshots_global}
\end{figure}
\end{center}

We now turn to the city map, in order to get a more accurate picture of the 
results so far. Fig.~\ref{tab:snapshots_global} displays the city as a 
square arrangement of blocks (see caption for details). Each ``pixel'' 
corresponds to a block, and the pixel color is associated to the corresponding 
scale on the left, which states the number of infected and the normalized number 
of removed per block. The snapshots capture the disease propagation from a 
single block located at the center of the map. A complete lockdown occurs at day 
100. 

The successive snapshots  display a seemingly symmetrical propagation pattern, 
shortly after the outbreak. However, ``secondary'' focuses appear around the 
main focus due to those long traveling individuals. Recall that we assumed that 
human mobility follows a Levy-flight distribution (see 
Section~\ref{sec:city}).\\

Notice that the lockdown implementation (from day 100 onwards) somehow 
``freezes'' the picture until the disease disappears (say, 50 days after). 
People continue to get infected within each block during the 
``quaratine'' period. Complementary health-care recommendations may be required 
for the disease control within each block. \\

\begin{figure}[!ht]
\centering
\includegraphics[width=0.7\columnwidth]{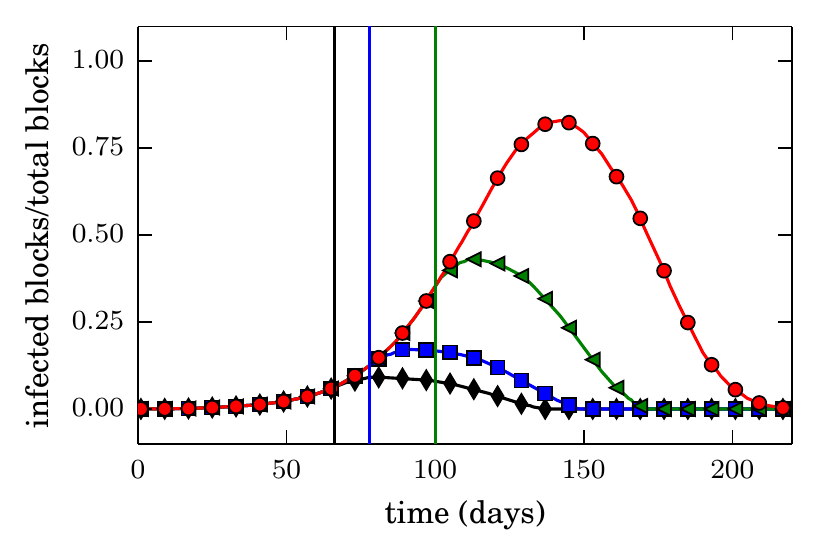}
\caption{\label{fig:focos_global} Normalized number of infected blocks as a 
function of time. The plot is normalized with respect to the total number of 
blocks ($14.4\,$k). Any block is classified as ``infected'' if at least one 
individual in the block is infected. We only consider people ``living in 
the block'' (say, those that sleep in there). The global lockdown is applied 
when the number of new infected equals to: \mydiamond{black}~5k, 
\protect\tikz\protect\draw[black,fill=blue] (0,0) rectangle (0.3cm,0.3cm); 
10k and \mytriangleleft{green} 30k new infected individuals. 
\protect\tikz\protect\draw[black,fill=red] (0,0) circle (.9ex); corresponds to 
the scenario of no lockdown at all. The infection rate remains constant along 
the simulation process and equals $\beta_0=0.75$. That is, there is no 
complementary health-care policies during the lockdown. The different lockdown 
implementation days are indicated by vertical lines.}
\end{figure}

Fig.~\ref{fig:focos_global} shows the number of ``infected blocks'',  
regardless of the number of infected people in the block (see caption for 
details). These curves quantify the infection map displayed in 
Fig.~\ref{tab:snapshots_global}, and resumes the effects of the full lockdown. 
\\

We may conclude that the GLs appears as a reliable strategy for 
avoiding the disease propagation. But the main drawback is that 
``non-infected'' blocks will enter the ``quarantine''. 
\ref{sec:global_complementarias} further shows the effects of complementary 
health-care policies.\\

\subsection{\label{sec:results_imperfect_lockdown}The imperfect global 
scenario (IGLs)}

In this case we model a situation in which the GLs cannot be implemented. We 
distinguish, however, two groups which cannot be kept confined: workers 
from essential activities (say, health care, food supply or public order 
services, etc.) or those who decide not to accept the confinement 
recommendation. The former are expected to follow complementary health-care 
recommendations, while the latter might not. For this reason, we 
will examine relaxed confinement conditions and infection rates 
reduction.\\

\begin{figure*}[!ht]
\centering
\subfloat[\scriptsize{Infected}\label{fig:infectados_imperfecta_5k}]
{\includegraphics[scale=0.8]{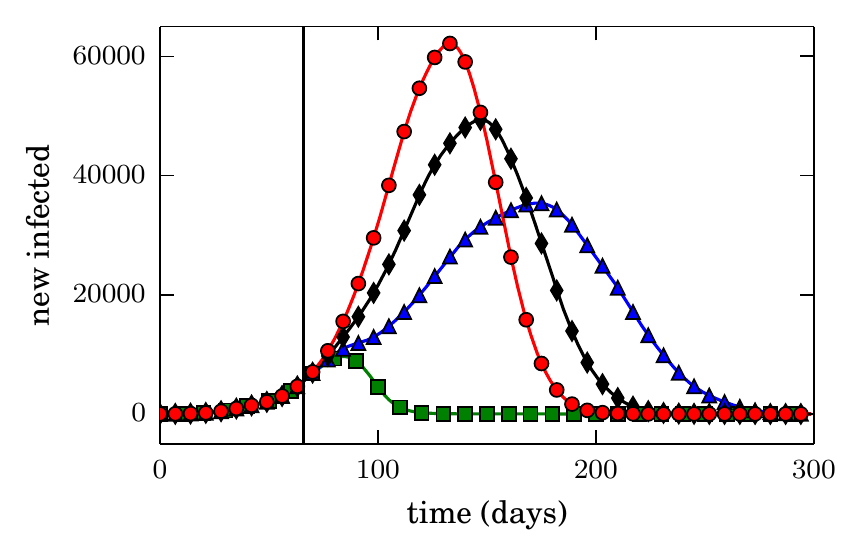}
}
\subfloat[\scriptsize{Removed}\label{fig:recuperados_imperfecta_5k}
] {\includegraphics[scale=0.8]{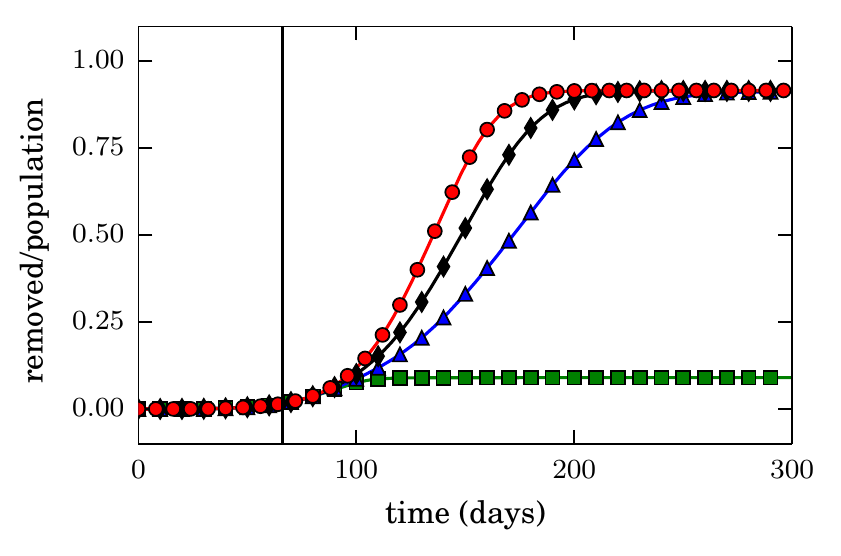}
}
\caption{\label{fig:infect_recup_imperfecta} (a) Number of new infected 
people and (b) normalized number of removed individuals as a function of 
time. The plot in (b) is normalized with respect to the total population in the 
city ($4.32\,$M). The IGLs is applied when the number of new 
infected equals to 5k (say, at day 66). From the lockdown 
implementation day, the percentage of individuals that move around is: 
\protect\tikz\protect\draw[black,fill=green] (0,0) rectangle (0.3cm,0.3cm); 
$0\%$, \mytriangleup{blue} $20\%$ and \mydiamond{black} $50\%$. 
\protect\tikz\protect\draw[black,fill=red] (0,0) circle (.9ex); corresponds to 
the scenario of no lockdown at all ($100\%$ of moving people). We consider a 
partial break of the mobility as the only heath-care policy. Thus, the 
infection rate equals $\beta_0=0.75$ all along the propagation process.} 
\end{figure*}

Fig.~\ref{fig:infectados_imperfecta_5k} shows how the infection curves change 
as the number of agents which do not accept the movement restriction increases  
(see caption for details). The confinement starts at the vertical line. Notice 
that the propagation stops dramatically for the complete confinement situation. 
But the possibility of stopping the outbreak vanishes if a fraction of people 
(as small as $20\%$)  still move around.  A quick inspection of 
Fig.~\ref{fig:recuperados_imperfecta_5k} confirms this point. \\

We also notice from Fig.~\ref{fig:recuperados_imperfecta_5k} 
that the total number of removed people at the end of the lockdown is the 
same for any IGLs. This appears to be in disagreement with the stochastic 
point of view, where the mobility reduction yields to the reduction in the 
probability of meeting people. We should remark that the meeting probability 
is somehow included in the infection rate $\beta$ within the SEIR model. Thus, 
for a complete picture of the IGLs, it is necessary to explore different 
values of $\beta$, as described below. \\

\begin{figure*}[!ht]
\centering
\subfloat[\scriptsize{}\label{fig:recup_vs_beta_imp_desde_66}]
{\includegraphics[scale=0.8]{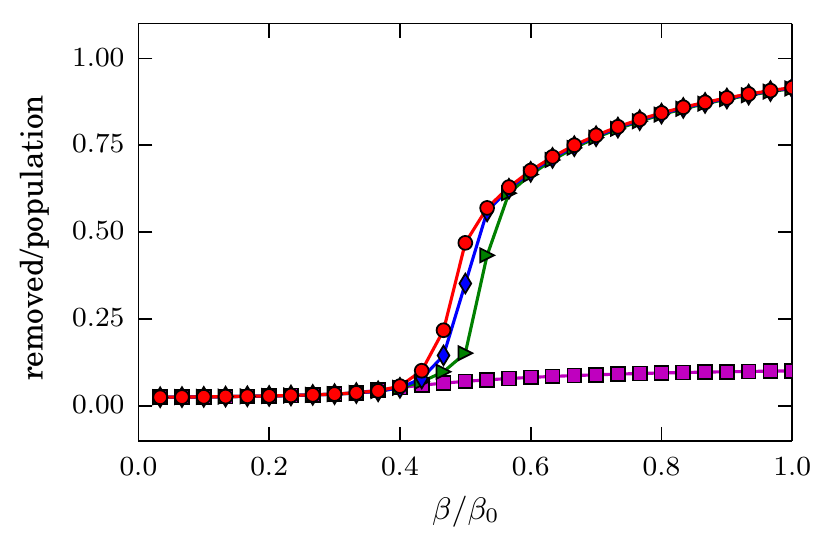}
}
\subfloat[\scriptsize{}\label{fig:recup_vs_beta_imp_desde_66}
]
{\includegraphics[scale=0.8]{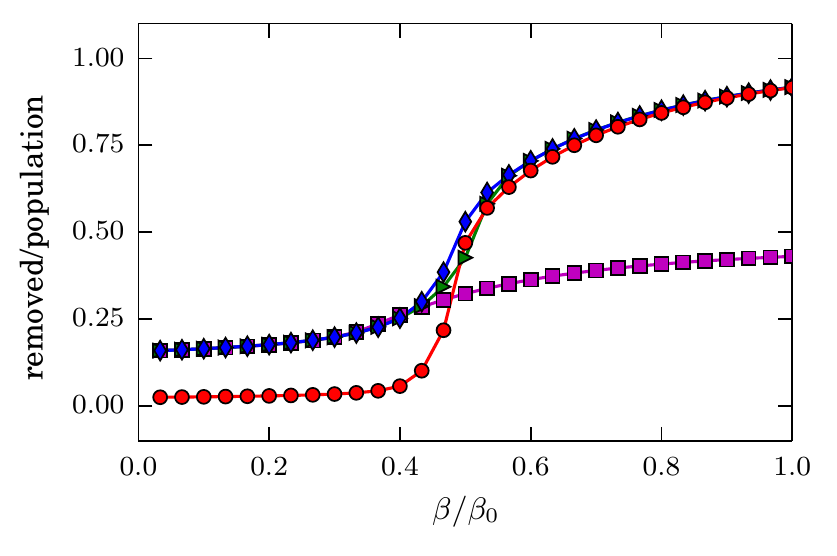}
}
\caption{\label{fig:recup_vs_beta_imp} Normalized number of removed people as
a function of the infection rate at the end of the epidemic. The plot is
normalized with respect to the total population in the city ($4.32\,$M). The
IGLs is applied when the number of new infected equals to: (a)
5k and (b) 30k. Since the lockdown implementation day, the percentage of
individuals that move around is: \protect\tikz\protect\draw[black,fill=magenta]
(0,0) rectangle (0.3cm,0.3cm); $0\%$, \mytriangleright{green} $20\%$ and
\mydiamond{blue} $50\%$. \protect\tikz\protect\draw[black,fill=red] (0,0)
circle (.9ex); corresponds to the scenario of no lockdown at all ($100\%$ of
moving people). Also, from the lockdown implementation day, the infection rate
change from $\beta_0=0.75$ to $\beta$. The city was composed by
$120\times120~$blocks placed on a square grid with 300 individuals per block.}
\end{figure*}

Fig.~\ref{fig:recup_vs_beta_imp} examines the number of removed individuals 
(\textit{i.e.} agents that have undergone the complete cycle of the illness, 
and have either reached a healthy state or have died) at the end of the epidemic 
in terms of the complementary health-care policies. Two startup 
days for the lockdown are shown  (see caption for details). We can see that 
the number of removed individuals experiences a dramatic change at 
$\beta/\beta_0\approx 0.5$ for the different moving people situations, 
regardless of the startup day of the lockdown.  This confirms the necessity of 
proper policies when these situations are expected to occur. \\

\begin{figure*}[!ht]
\centering
\subfloat[$0\%$ of mobility\label{fig:recup_0_porc_imp}]{
\includegraphics[scale=0.35]{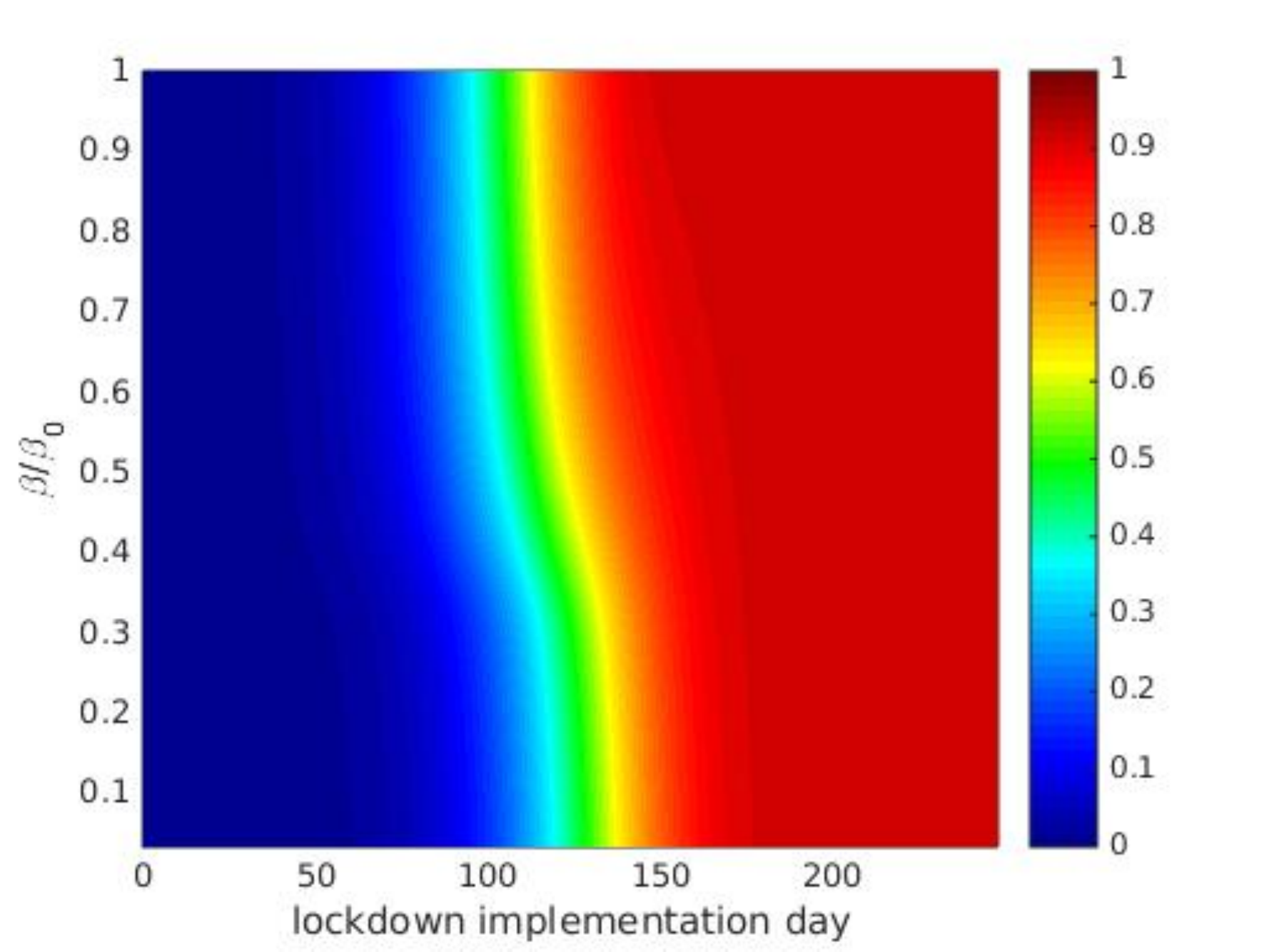}
}
\hspace{-1mm}
\subfloat[$50\%$ of mobility\label{fig:recup_50_porc_imp}]{
\includegraphics[scale=0.35]{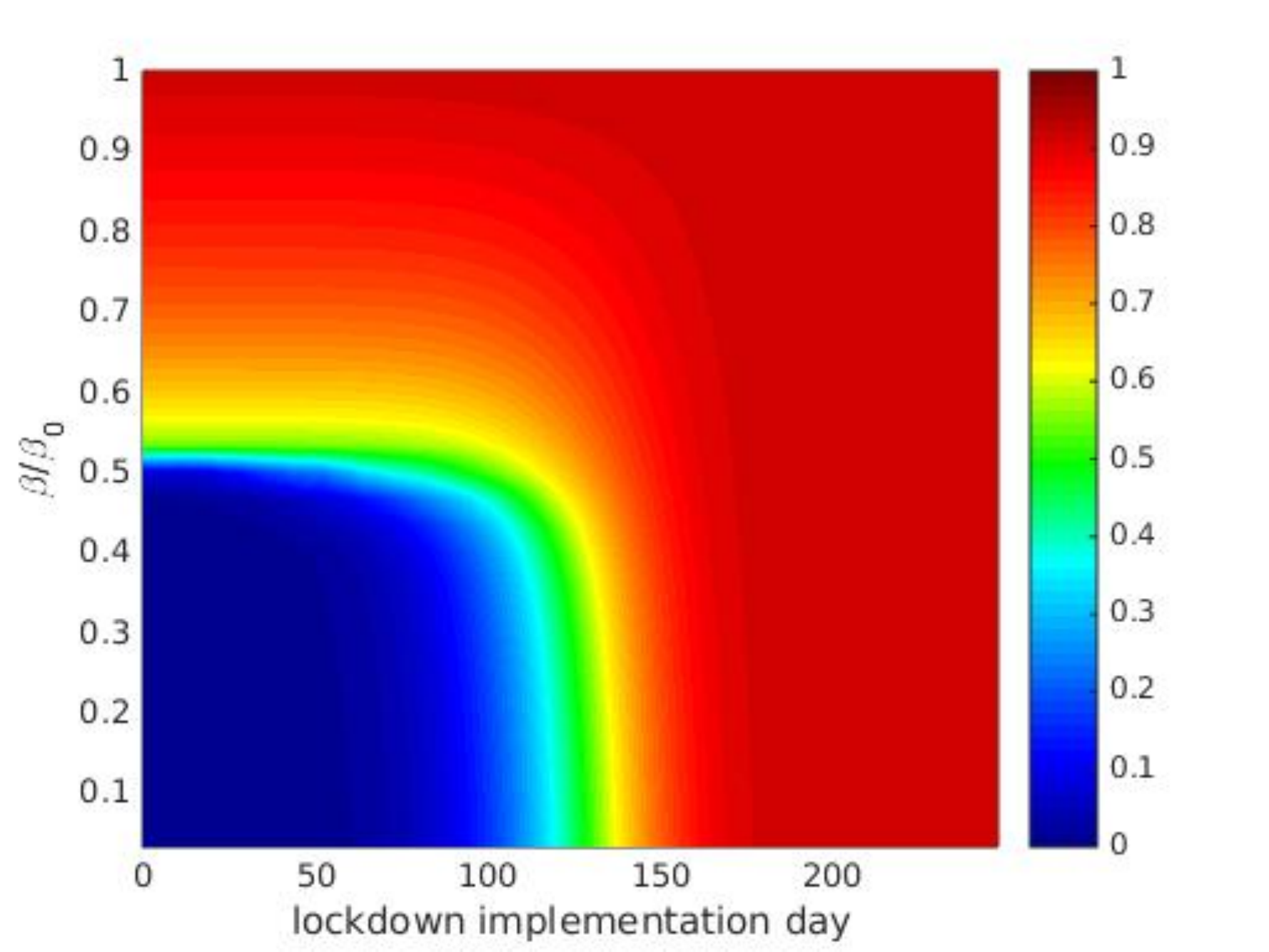}
}
\caption{\label{fig:mapa_recuperados_imp} Normalized number of removed 
individuals (see scale on the right) as a function of the lockdown 
implementation day and the infection rate ($\beta/\beta_0$). The normalization 
was done taking into account the city population. The infection rate changes 
from $\beta_0=0.75$ to $\beta$ since the lockdown implementation day.} 
\end{figure*}

We improved on the above results by displaying in 
Fig.~\ref{fig:mapa_recuperados_imp} the contour maps for different mobility 
situations, as a function of the lockdown date. Notice that 
Fig.~\ref{fig:recup_vs_beta_imp} corresponds to date 66 in both contour maps 
(see caption in Fig.~\ref{fig:recup_vs_beta_imp}). \\

Fig.~\ref{fig:mapa_recuperados_imp} expresses the fact that whenever the human 
mobility is suppressed, the number of removed individuals depends (almost 
exclusively) on the timing of the lockdown implementation. That is, the 
earlier the lockdown, the lower the number of removed. Besides, if mobility 
cannot be suppressed (say, $50\%$ of the people still move around), then 
complementary health-care policies should be heavily implemented, in order to 
avoid a massive contagion. This appears as an essential issue for late 
lockdowns.  \\ 

We close this section with the following conclusion: a complete mobility 
suppression appears as the most effective way of reducing the number of 
casualties. Essential workers following strict health-care recommendations 
(say, masks, behavioral protocols, etc.) that move around, however, will not 
spread the disease to uncontrolled levels. But, a small fraction of people 
moving around out of protocol can spoil the mitigation efforts. \\

\subsection{\label{sec:results_local_lockdown}Local lockdown scenario (LLs)}

The local lockdown means that people remain confined within the block where 
he (she) lives, depending on the infection level of their block. That is, those 
blocks surpassing a certain ``threshold'' of infected people are immediately 
isolated, while the others remain ``open''. As in the case of the GLs, we assume 
that people from an isolated block may still get in contact within this block. 
We consider the mobility break as the only heath policy.\\

\begin{figure*}[!ht]
\centering
\subfloat[\scriptsize{Infected}\label{fig:infec_rl_3_0_local}]
{\includegraphics[scale=0.8]{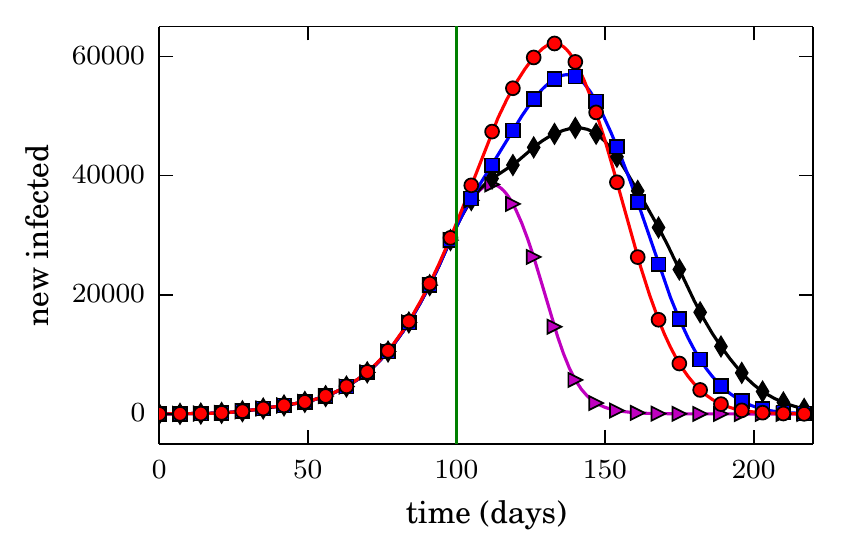}
}
\subfloat[\scriptsize{Removed}\label{fig:recup_rl_3_0_local}
] {\includegraphics[scale=0.8]{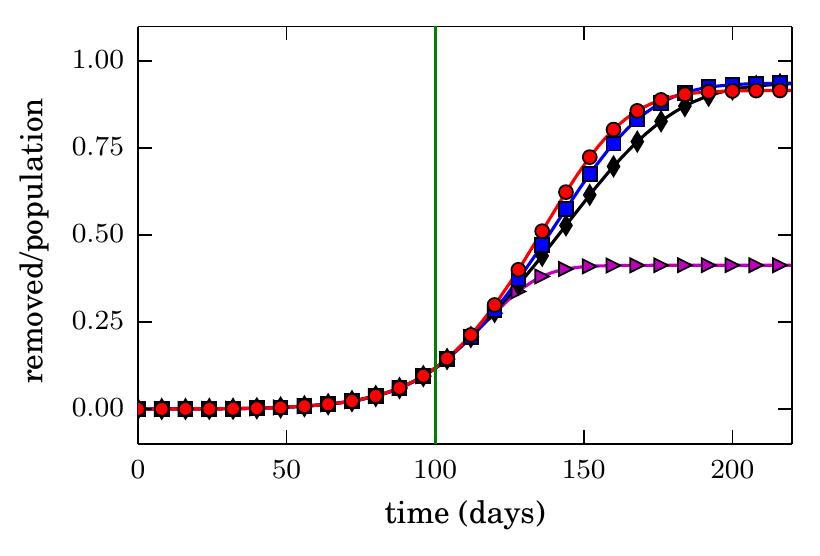}
}
\caption{\label{fig:infec_tiempo_local} (a) Number of new infected people and 
(b) normalized number of removed individuals as a function of time. The plot 
in (b) is normalized with respect to the total population in the city 
($4.32\,$M). The LLs is applied when the number of new infected 
equals to 30k (indicate by the vertical line, at day 100). The locked blocks are 
those that exceed the following thresholds: \mytriangleright{magenta} 0, 
\mydiamond{black} 5 and \protect\tikz\protect\draw[black,fill=blue] (0,0) 
rectangle (0.3cm,0.3cm); 10 individuals (see text for details). 
\protect\tikz\protect\draw[black,fill=red] (0,0) circle (.9ex); corresponds to 
the scenario of no lockdown at all. We consider the isolation of the infected 
blocks as the only heath-care policy. Thus, the infection rate equals  
$\beta_0=0.75$ all along the propagation process.} 
\end{figure*}

The local lockdown relies on the idea that the early detection of the infected 
can prevent the disease from spreading around the city. This idea presumes 
insignificant or null detection flaws. But in practice, a few infected 
individuals may not be detected due to wrong testings, or do not present 
symptoms at all. Our concern is on these situations. We will assume that a 
small fraction of infected people per block cannot actually be detected. This 
means that the lockdown occurs after surpassing a ``threshold'' of infected 
people (although not detected). \\

Fig.~\ref{fig:infec_tiempo_local} represent the usual infection and passed over 
curves as a function of time, respectively (see caption for details). A quick 
inspection of these curves show that the disease decays almost immediately if 
the lockdown process occurs just after the first infected appeared in the 
block. Otherwise, the number of infected people continues increasing until the 
day 150 (approximately). \\

Fig.~\ref{fig:infec_tiempo_local} can be compared to its counterpart in the 
context of the IGLs, that is, Fig.~\ref{fig:infectados_imperfecta_5k}. The time 
scales in both situations appear quite similar, although the curves in 
Fig.~\ref{fig:infec_rl_3_0_local} do not extend further than 200 days. \\

\begin{figure*}[!ht]
\centering
\subfloat[\scriptsize{Infected threshold 0}\label{fig:snap_0_umbral_local}]{
\includegraphics[scale=0.35]{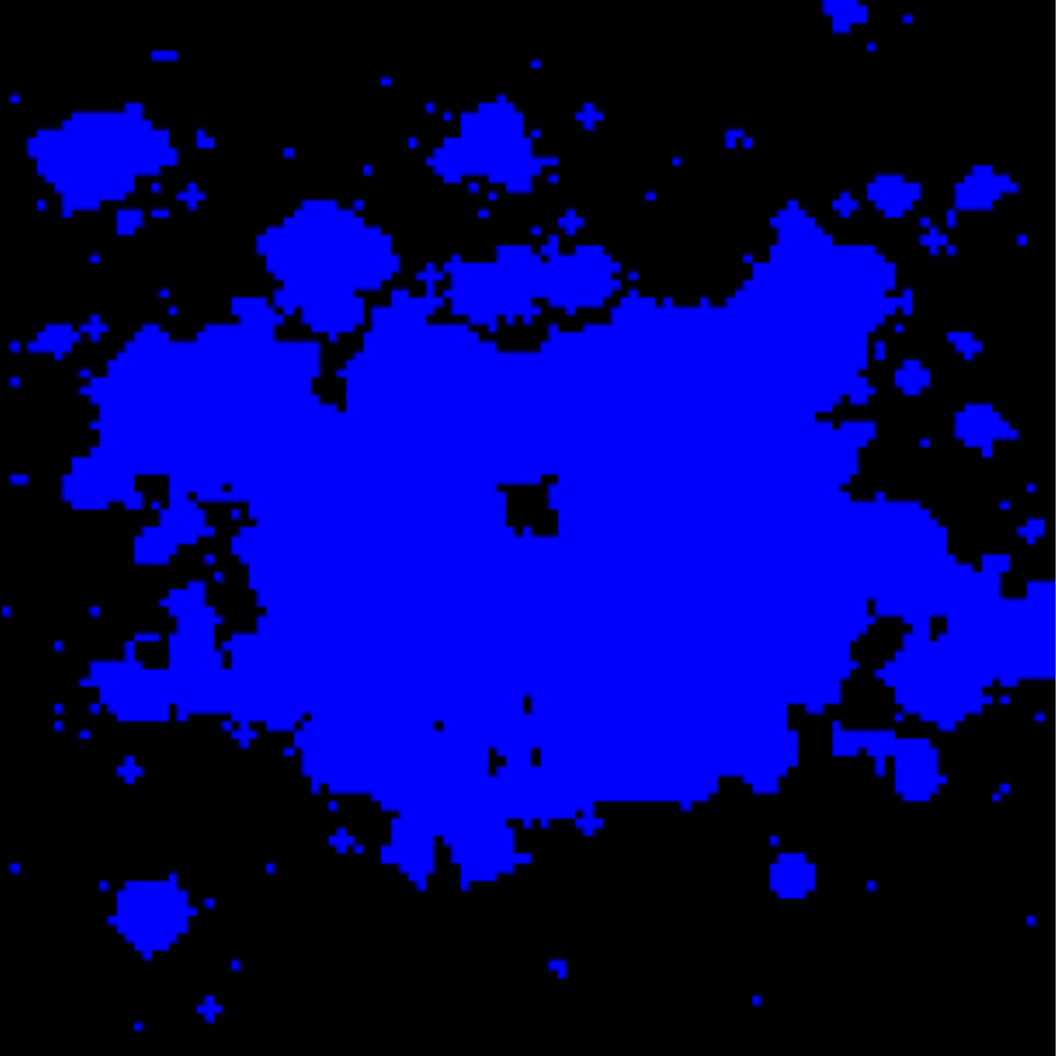}
}
\subfloat[\scriptsize{Infected threshold 5}\label{fig:snap_5_umbral_local}]{
\includegraphics[scale=0.35]{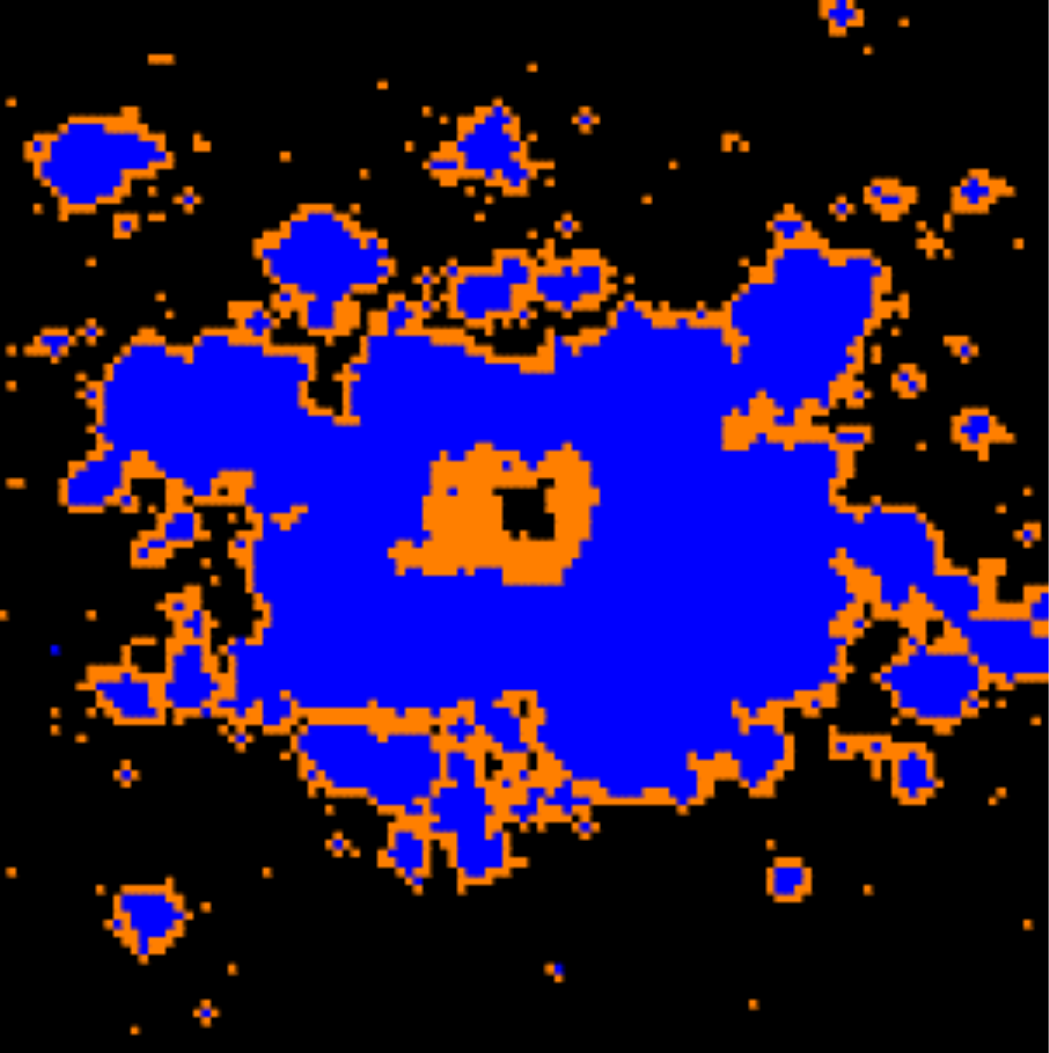}
}
\subfloat[\scriptsize{Infected threshold 20}\label{fig:snap_20_umbral_local}]{
\includegraphics[scale=0.35]{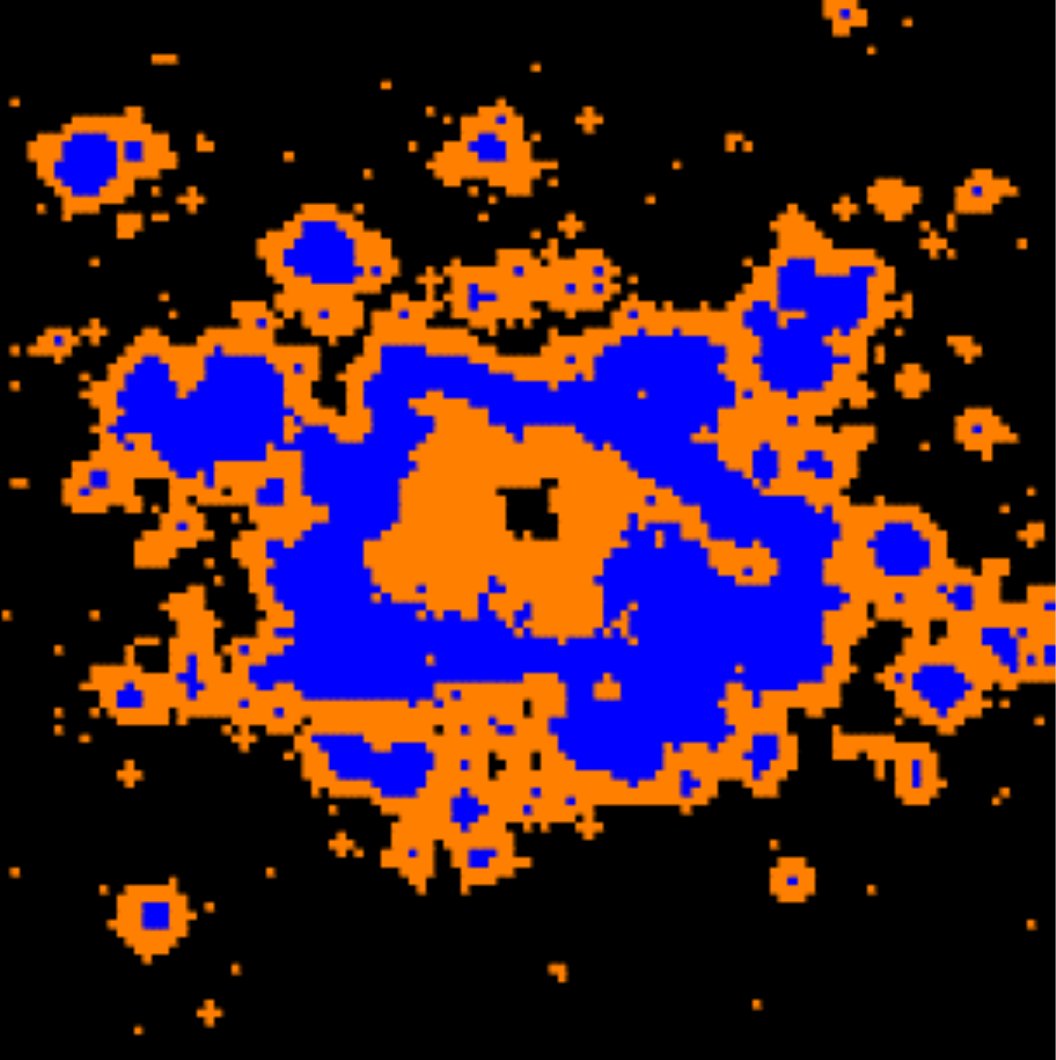}
}
\caption{\label{fig:snapshots_local} Spatial distribution of the blocks 
according to their infected state at the lockdown implementation day (100). 
Those blocks without infected individuals are represented in black, while those 
blocks with a number of infected people lower (greater) than the threshold 
(see legend) are represent in orange (blue). Blue blocks were isolated 
from the rest, while black and orange blocks were ``opened''. The infection 
rate remains constant along the simulation process and equals $\beta_0=0.75$. 
The city was composed by $120\times120~$blocks placed on a square grid with 300 
individuals per block.} 
\end{figure*}

The ``threshold'' of non-detected people is responsible for the time lapse 
between the lockdown day and the end of the disease. This can be confirmed 
through Fig.~\ref{fig:snapshots_local}, that shows the contour maps for the 
infected blocks (see caption for details). Notice that the non-detected blocks 
(say, the orange ones) become more relevant as the ``threshold'' level 
increases. \\

\begin{figure}[!ht]
\centering
\includegraphics[width=0.6\columnwidth]{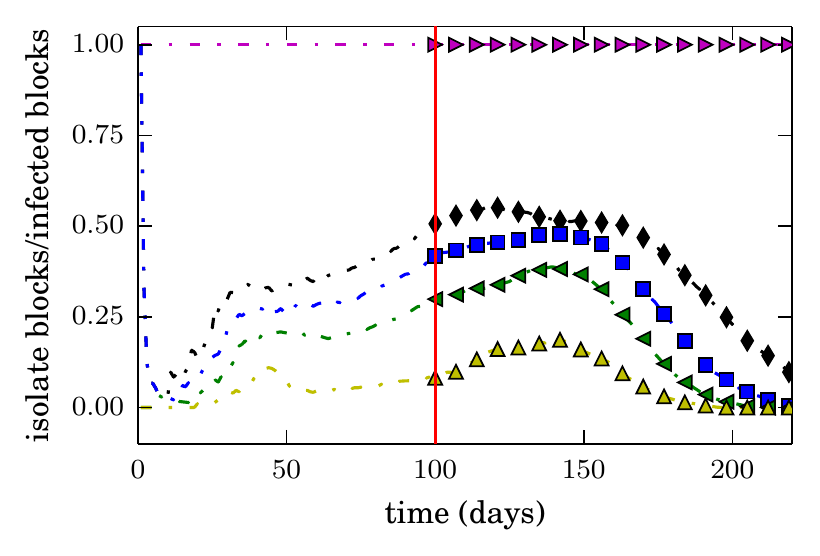}
\caption{\label{fig:eficiencia_local} Normalized number of isolate blocks 
with respect to the total number of infected blocks as a function of time. The 
LLs is applied when the number of new infected equals to 30k 
(indicated by the vertical line, at day 100). The locked blocks are 
those that exceed the following thresholds: \mytriangleright{magenta} 
0, \mydiamond{black} 5, \protect\tikz\protect\draw[black,fill=blue] (0,0) 
rectangle (0.3cm,0.3cm); 10, \mytriangleleft{green} 20 and \mytriangleup{yellow} 
40 individuals (see text for details). The dashed lines corresponds to 
the behavior of both magnitudes before the lockdown implementation. The 
infection rate remains constant along the simulation process and 
equals $\beta_0=0.75$.}
\end{figure}

Fig.~\ref{fig:eficiencia_local} exhibits the fraction of isolated blocks with 
respect to those blocks attaining at least an infected individual. We can 
observe that half of the infected blocks are actually not detectable for a 
threshold as low as 5 individuals per block. This is a strong warning on the 
effectivity of the LLs. Public health officers will lock down as many 
blocks as detected, but the undetected will actually continue the propagation. 
\\

\begin{figure}[!ht]
\centering
\includegraphics[width=0.7\columnwidth]{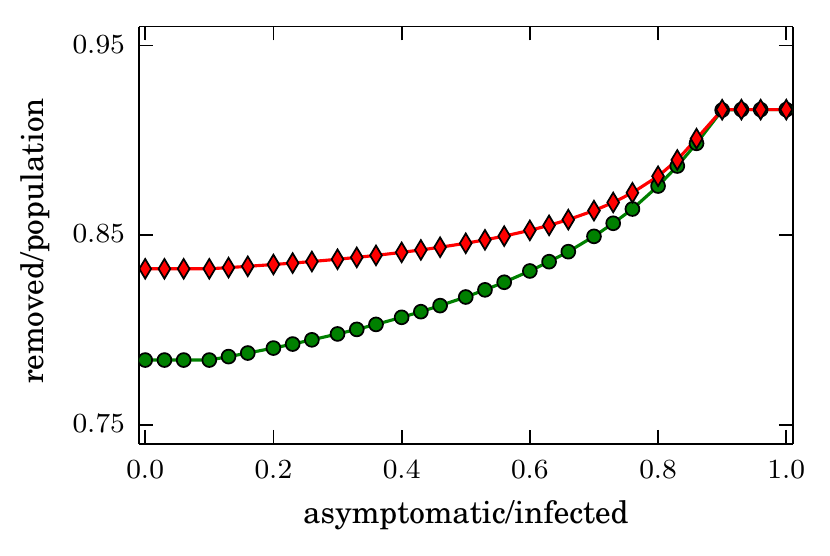}
\caption{\label{fig:plot_asint_vs_porc} Normalized number of removed 
individuals (at the end of the epidemic) as a function of the percentage of 
asymptomatic individuals. The normalization was done taking into account the 
city population. The LLs started at the day 100 after the first 
infected was detected. The infection threshold is five individuals. From the 
lockdown implementation day, the infection rate switches from $\beta_0=0.75$ 
to $\beta=0.075$ in \protect\tikz\protect\draw[black,fill=green] (0,0) circle 
(.9ex);, while remains constant ($\beta_0$) in \mydiamond{red}.}
\end{figure}

Whatever the efforts to detect infected, it seems that $40-45\%$ of the 
infected individuals do not experience noticeable symptoms 
\cite{Asymptomatic}. We introduced this phenomenon into our simulations. 
Fig.~\ref{fig:plot_asint_vs_porc} shows the overall removed people for an 
increasing number of ``non-symptomatic'' individuals. This confirms once more 
the lack of effectivity  of the local lockdown if no other policy is 
established. \\
 
A thoughtful policy should include either strategic testings and backward 
tracing of the infected, from our point of view. We simulated this policy by 
tracing back any infected people to where she (he) belonged before being 
detected. The procedure in a nutshell is as follows:

\begin{itemize}
 \item Test a random block. If infected, trace back \textit{all} the individuals 
to the block they visited before.
 \item Test \textit{all} the blocks recognized as visited immediately before. 
 \item Lock down any of the above if infected. \\
\end{itemize}

The results from these simulations are shown in \ref{sec:testing_tracing}. The 
testing procedure exhibits a noticeable efficiency (say, a noticeable decay in 
the number of removed people) if at least $80\%$ of the blocks can be tested. 
The back-tracing procedure further reduces this fraction significantly. 
Other complementary health-care policies (like masks, distancing, etc.) can 
also improve significantly the number of removed individuals (see 
\ref{sec:testing_tracing} for details). In summary, the simulation results 
confirm our intuition on the effectivity of the back-tracing methodology. \\

We conclude from this Section that the effectivity of any LLs will 
strongly depend on breaking off the mobility of infected people. If this fails 
(because of asymptomatic or  unreachable people), the  disease will spread 
dramatically throughout the city. The strategic testing and back-tracing of 
the infected should be considered as an essential tool for the disease 
mitigation. \\

\section{\label{sec:discussion}Analysis of the effects of the different 
strategies}

In this Section we discuss the performance of the GLs, IGLs and 
LLs. We limit our analysis to the following 
points  

\begin{itemize}
 \item[(a)] The performance of the lockdown is actually associated to the 
mitigation of the disease. Smooth infection curves are preferred in 
order to avoid stressing the medical care system. 

\item[(b)] Lockdowns seriously damage the economy. The less disturbing and 
shorter lasting actions on non-infected people are therefore preferred.  \\

\end{itemize}

We propose a merit function in order to rate the performance of the 
different lockdown strategies above mentioned, with respect to conditions (a) 
and (b). We will consider the fraction of the new infected 
people at any time $i(t)$ (or $i_n$ at step $n$ of the simulation) and the 
mobility $\mu(t)$ as the most relevant quantities for building 
the merit function (see below for the precise definitions). Thus, the merit 
function will be expressed as $C=C(i,\mu)$. \\ 

Notice from Section \ref{sec:results} that the maximum number of new 
infected people is quite different for the examined scenarios (see, for 
example, Figs.~\ref{fig:infect_recup_imperfecta} and 
\ref{fig:infec_tiempo_local}). Our merit function will consider the new 
infected people ($i$) normalized with respect to the maximum number of 
new infected when no lockdown is carried out. Accordingly, we will 
consider the mobility fraction ($\mu$) as the amount of traveling people 
normalized with respect to the traveling people before the lockdown.  \\

The topic (a) concerns the infected people. The successful lockdown will
 mitigate the overall number of infected people. Our first proposal would 
be to rate the performance as the cumulative value of $i$ along the 
lockdown period. This approach, however, does not consider the stressing of
 the medical care system. For instance, it does not make any difference 
 between sharp infection curves and smooth ones, provided that the total 
number of infected people are the same. We can therefore improve the proposed 
function by cumulating the fractions $i^\alpha$, for $\alpha>1$. 
The coefficient $\alpha$ introduces a penalty to the sharp maximum 
(see below).  \\

The topic (b) concerns the non-infected people. The lockdown breaks the 
routine of the traveling fraction of people $\mu\,(1-i)$, and consequently, 
the economic activity. We propose rating the performance of the non-infected 
motion as the aggregate of a linear function of $\mu\,(1-i)$. \\

We express our merit function as follows\\

\begin{equation}
 C=\displaystyle\sum_{n=1}^N\bigg[
 \overbrace{i_n^\alpha}^{\mathrm{(a)}} + \overbrace{
 A+B\,\mu_n\cdot(1-i_n)}^{\mathrm{(b)}}\bigg]
\label{eqn:cost_function_0}
\end{equation}

\noindent where $n=1...N$ stands for the day of the lockdown. 
The term (a) refers to the medical care cost, and the term 
(b) refers to the economical cost.   \\

Notice that in regular working days $i=0$ and $\mu=1$. This 
yields a daily economical cost equal to $A+B$, according to 
(\ref{eqn:cost_function_0}). We rate this cost as the null cost ($C=0$) for 
practical reasons. Thus, we set $A+B=0$ to hold this condition. The cost 
function then reads  \\

\begin{equation}
 C=\displaystyle\sum_{n=1}^N\bigg[
 \overbrace{i_n^\alpha}^{\mathrm{(a)}} + \overbrace{
 A\,[1-\mu_n\cdot(1-i_n)}^{\mathrm{(b)}}]\bigg]
\label{eqn:cost_function}
\end{equation}

This expression shows that an increase in the number of new 
infected people (although keeping $\mu=1$) yields an increase in the medical 
cost (a) and the economical cost (b). The implementation of a strict 
lockdown ($\mu\rightarrow 0$) further increases the economical cost (b) to 
its maximum value.  The intermediate situations will be more 
or less costly according to the balance between the medical care cost (a)  
and the economical cost(b) in expression (\ref{eqn:cost_function}). The 
parameter $A$ is the decisive parameter in the balance between the medical 
care, and economical cost. \\

We stress that the proposed function (\ref{eqn:cost_function}) is limited 
to items (a) and (b), while other presumably important arguments could 
have been left aside for simplicity. We also fixed $\alpha=2$ for 
practical reasons, but we checked that $C$ behaves qualitatively the 
same for other values $\alpha>1$ (not shown). \\

The parameter $A$ is actually the only free parameter in our cost function. 
It stands as a weighting factor for the economical cost. We will discuss 
the behavior of $C$ for the lockdown strategies appearing in Section 
\ref{sec:results} while varying the values of $A$.  \\

\begin{figure*}[!ht]
\centering
\subfloat[\label{fig:plot_costo_imperfecta_discrimindo}]{
\includegraphics[scale=0.8]{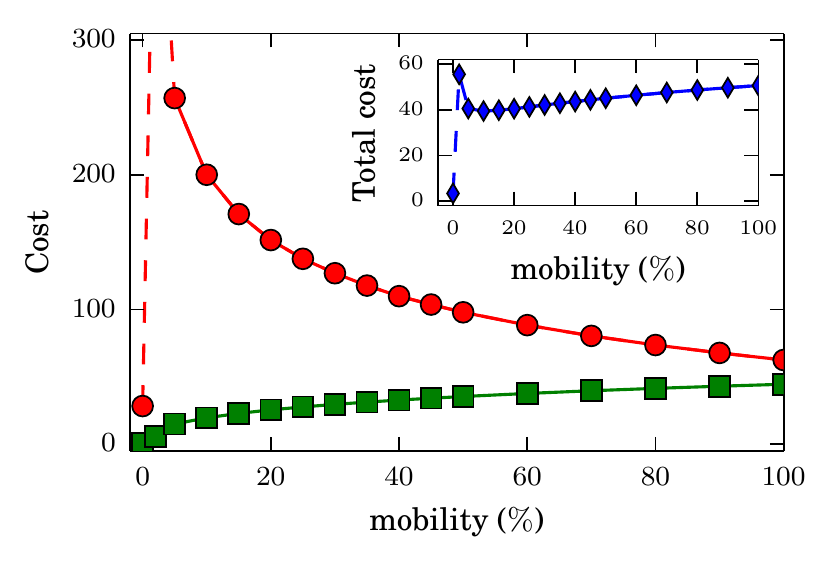}}
\subfloat[\label{fig:plot_costo_local_discriminado}]{
\includegraphics[scale=0.8]{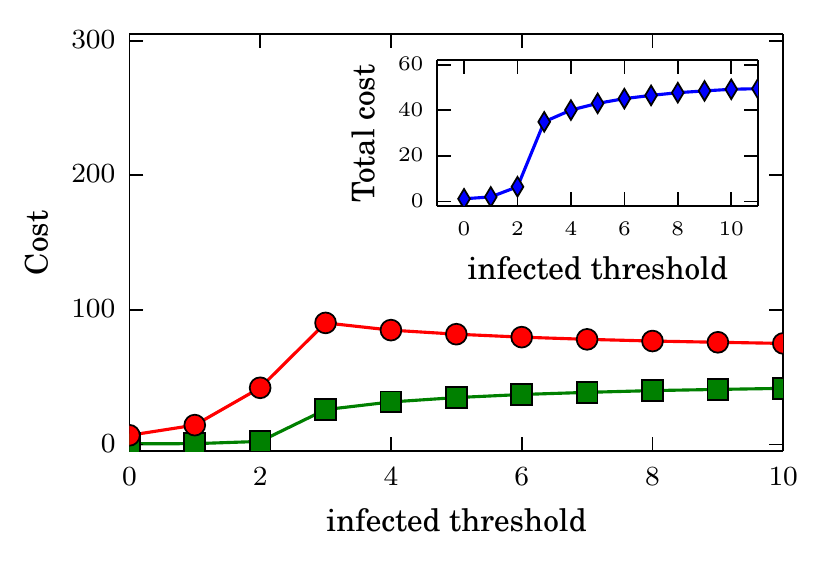}}
\caption{\label{fig:plot_costos_discriminados}  Medical care cost 
(\protect\tikz\protect\draw[black,fill=green] (0,0) 
rectangle (0.3cm,0.3cm);) and economical cost 
(\protect\tikz\protect\draw[black,fill=red] (0,0) circle 
(.9ex);) for the scenarios analyzed in Section \ref{sec:results}. The inset 
shows the total cost $C$ assuming $A=0.1$. (a) Implementation of the GLs and 
IGLs vs. $\mu$. The dashed lines means that the economical cost for mobilities 
between $0\%$ and $5\%$ are greater than 300. (b) Implementation of the LLs 
vs. the infection threshold. In both cases, the lockdown is applied when the 
number of new infected equals to 5k. The infection rate remains constant along 
the simulation process and equals $\beta_0=0.75$.} 
\end{figure*}

Let us first examine the medical care cost (a) and economical cost (b) 
separately. Fig.~\ref{fig:plot_costos_discriminados} plots these costs for the 
situations shown in Figs.~\ref{fig:infect_recup_imperfecta} and 
\ref{fig:infec_tiempo_local}, respectively (see caption for details). 
The varying parameter is different on each plot, that is, the 
horizontal axis corresponds to the mobility in 
Fig.~\ref{fig:plot_costo_imperfecta_discrimindo} and to the threshold level in
Fig.~\ref{fig:plot_costo_local_discriminado}. The horizontal scale in 
Fig.~\ref{fig:plot_costo_imperfecta_discrimindo} runs from the most strict 
situation at the origin ($\mu=0$) to the normal moving situation ($\mu=1$). 
Analogously, the horizontal scale in 
Fig.~\ref{fig:plot_costo_local_discriminado} runs from the most early detection 
at the origin to a detection level of 10 individuals. We present, however, both 
plots together in order to visualize the behavior of the GLs, IGLs and LLs as 
the lockdown becomes more and more relaxed. 
Fig.~\ref{fig:plot_costos_discriminados} resumes, indeed, the costs as a 
function of the lockdown strictness.     \\

We notice from Fig.~\ref{fig:plot_costos_discriminados} that 
the medical cost, although different, shows a quite similar behavior in 
the IGLs and the LLs (see green squares in there). The economical costs, 
however, differ from each other when the lockdowns are very strict. The mutual 
differences vanish for the relaxed situations, no matter if the lockdown is 
global 
or local. \\

The dramatic increase of the economical cost (for the strict 
global lockdowns) is a matter of concern. 
Fig.~\ref{fig:plot_costo_imperfecta_discrimindo} 
reports this phenomenon for small mobility values (IGLs), although not for the 
null mobility situation (GLs). The null mobility situation means that the 
disease remains confined to each block. But if a few agents avoid the 
lockdown, they can additionally spread the disease to other non-infected 
blocks. Thus, the perfect lockdown situation and the imperfect 
situation are quite different ones. This can be verified by observing 
Fig.~\ref{fig:infect_recup_imperfecta}. The infection curve smoothens and 
widens as the mobility switches from 0\% to 20\%. For a mobility fraction of 
50\% the curve narrows back again.  \\

Either Fig.~\ref{fig:infectados_imperfecta_5k} and 
Fig.~\ref{fig:plot_costo_imperfecta_discrimindo} point out that small 
mobility values induce a slow spreading dynamics (\textit{i.e.}
no massive propagation). This does not stress the medical care system, but 
yields long disruptions of the working routines. Notice from the cost 
expression (\ref{eqn:cost_function}) that the term (b) is linear to $N$ if 
$\mu\approx 0$. Thus, the long lasting lockdowns are responsible for the  
 increase in the economical costs. \\

The LLs isolates the infected blocks only. The mobility is 
locally suppressed, and the whole scene undergoes a mixture of blocks with 
full mobility and blocks that are isolated. This is a dynamical process where 
the blocks become isolated and are opened back again. The net result is that 
the  
overall mobility is not significantly reduced along the lockdown, and 
therefore, the spreading dynamic does not ``slow down'' completely, as 
already noticed in Fig.~\ref{fig:infec_rl_3_0_local}. Recall that the curves 
therein do not extend further than 200 days, no matter the detection 
threshold. This prevents the dramatic increase in the economic costs, as shown 
in Fig.~\ref{fig:plot_costo_local_discriminado}. \\

\begin{figure*}[!ht]
\centering
\subfloat[\label{fig:plot_costo_imperfecta}]{
\includegraphics[scale=0.8]{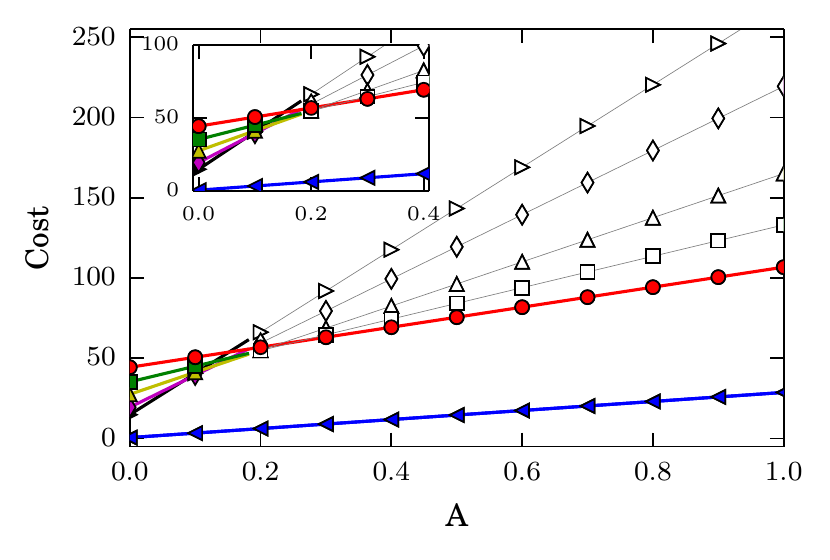}}
\subfloat[\label{fig:plot_costo_local}]{
\includegraphics[scale=0.8]{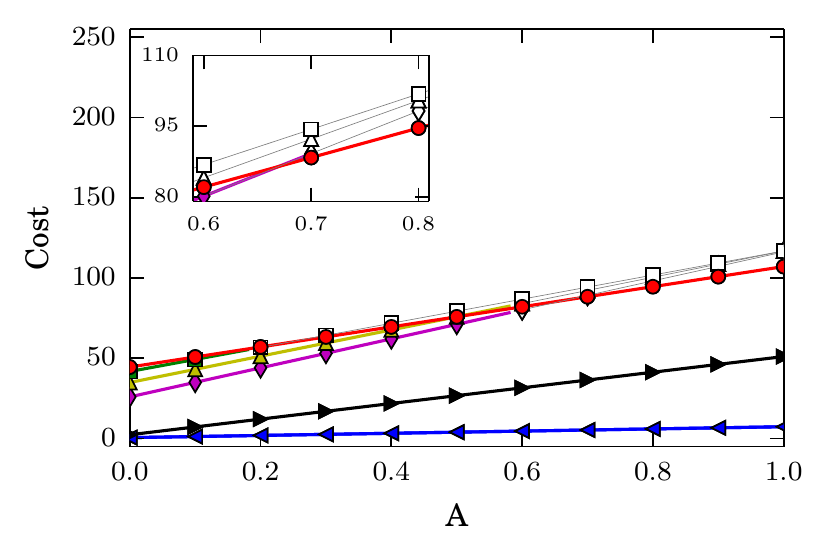}}
\caption{\label{fig:plot_costos} (a) Cost function $C$ as a function of 
the weighting parameter $A$ for the GLs and IGLs. From the lockdown 
implementation day, the percentage of individuals that move around is: 
\mytriangleleft{blue} $0\%$ (GLs), \mytriangleright{black} 
$5\%$, \mydiamond{magenta} $10\%$, \mytriangleup{yellow} $25\%$ and 
\protect\tikz\protect\draw[black,fill=green] (0,0) rectangle (0.3cm,0.3cm); 
$50\%$. (b) Cost function $C$ as a function of the weighting parameter $A$ for 
the LLs. The locked blocks are those that exceed the 
following thresholds: \mytriangleleft{blue} 0, \mytriangleright{black} 
2, \mydiamond{magenta} 3, \mytriangleup{yellow} 5 
and \protect\tikz\protect\draw[black,fill=green] (0,0) rectangle (0.3cm,0.3cm); 
10 individuals. In both cases, \protect\tikz\protect\draw[black,fill=red] (0,0) 
circle (.9ex); corresponds to the scenario of no lockdown at all. Also, the 
lockdown is applied when the number of new infected equals to 5k. The infection 
rate remains constant along the simulation process and equals $\beta_0=0.75$.} 
\end{figure*}

We turn now to the discussion of the different lockdown strategies  
performance. Fig.~\ref{fig:plot_costos} shows the total cost $C$ as a function 
of the weighting factor $A$ (see caption for details). Notice that although 
$A$ can be chosen freely, it seems unrealistic to allow values yielding to 
costs beyond the cost of no-lockdown at all. We will consider the unlocked 
situation as the bounding value to $C$, as shown in red in 
Fig.~\ref{fig:plot_costos} (see caption for details). \\

Fig.~\ref{fig:plot_costos} reports the minimum cost situation in blue color 
(\mytriangleleft{blue}) for the GLs and LLs. These correspond to 
either the most strict global lockdown ($\mu=0$) or the most early detection 
for the local lockdown (say, the null threshold). The latter attains a better 
performance with respect to the former for all the explored values of $A$.  \\

The economical cost (b) is responsible for the slope of the curves in 
Fig.~\ref{fig:plot_costos}. The almost flat slope observed for the null 
threshold means that the daily cost for this situation is negligible. 
This is in agreement with an early detection, where the fraction of 
infected people is quite small ($i\approx 0$) and most of the non-infected 
people are allowed to move around ($\mu\approx 1$). The reader can check 
that this yields to vanishing values for the term (b) in our cost 
model (\ref{eqn:cost_function}). \\

We further notice that the curves' slope increases for increasing 
detection thresholds in Fig.~\ref{fig:plot_costo_local}. This occurs 
because of the increase in the daily economical cost, although the lockdown 
period remains close to $N\approx 200\,$days. Interestingly, the lockdown 
curves in Fig.~\ref{fig:plot_costo_local} meet the no-lockdown curve for 
thresholds surpassing 2-3 infected people. This makes the lockdown curves 
only valid for small values of $A$ (say, below 0.7).   \\

Recall that the IGLs experiences a dramatic increment of the economical cost 
for small mobility fractions (see 
Fig.~\ref{fig:plot_costo_imperfecta_discrimindo}). 
As a result, the lockdown curves always meet the no-lockdown curve at some point 
(except for $\mu=0$), as shown in Fig.~\ref{fig:plot_costo_imperfecta}. This is 
quite a difference with respect to the local strategy, since the lockdown 
curves are always limited to small values of $A$ (except for $\mu=0$). \\

Let us close the discussion with the following comments. We showed that the 
most strict implementation of either the global or local lockdown leads to the 
optimum performance (although the local one is preferred, as discussed 
above). But we noticed that there is some space left for partially 
effective lockdowns, if the most strict conditions are not attainable. 
The degree of effectiveness depends on the balance between the 
medical care costs and the economical costs, within this model.  \\

\section{\label{sec:conclusions}Conclusions}

This work concerns with the effect of human mobility in the 
context of a model for the spatio-temporal evolution of the COVID-19 outbreak. 
People move according to the Levy distribution and get into contact with 
each other during their daily routine. The lockdowns prevent these contacts 
from occurring, and thus, mitigate the propagation of the disease. Our 
investigation studies different lockdown scenarios and performs a careful 
evaluation of their effectiveness. We draw some  recommendations for the 
better performance of the lockdown.  \\

We assumed a SEIR compartmental model (with constant 
infection rates) for the inhabitants of a block. Each block was considered as 
a node within a square network of $120\times120$ nodes. People were allowed to 
travel twice a day between blocks. The initial conditions for the simulations
considered 20 infected individuals located at the central block of the 
network, while the rest of the individuals were assumed to be in the 
susceptible state.\\

We focused on three scenarios: the full lockdown of all the 
blocks (perfect scenario), the partial lockdown of all the blocks (imperfect 
scenario), and the lockdown of only the infected blocks (local lockdown). We 
sustained the lockdown until the disease propagation was (almost) over. 
\\

We first noticed that the success of any control action 
depends strongly on canceling the mobility around the city. But a small number 
of individuals may spoil the effectiveness of these actions if they do not 
follow the confinement recommendations. This is, in our opinion, the major 
risk when implementing any lockdown strategy. \\

We further built a cost function to rate the three strategies. We arrived to 
the conclusion that full lockdowns, or, the (very) early detection and 
isolation strategy are the most effective ones. The local isolation strategy 
is preferred, though, since it appears as the less costly in the context of 
our model. \\

It is important to emphasize that strict lockdown policies 
also allow for short periods of isolation. More relaxed lockdowns are less 
costly daily, but cumulate large costs after an extended period of time.\\  

The full lockdown looses effectivenesses if the mobility is 
not completely canceled, as already mentioned. Our model shows that local 
lockdowns can still be quite effective even if a small number of infected 
people is not detected (and isolated). But the ultimate decision on whether to 
choose a global or local lockdown on a specific circumstance will depend on 
the right balance between the medical care cost and the economical cost due to 
routine disruptions. We will leave this discussion open to future research. 
\\

\section*{Acknowledgments}
This work was supported by the National Scientific and Technical 
Research Council (spanish: Consejo Nacional de Investigaciones Cient\'\i ficas 
y T\'ecnicas - CONICET, Argentina) and grant Programaci\'on Cient\'\i fica 2018 
(UBACYT) Number 20020170100628BA. G. Frank thanks Universidad Tecnol\'ogica 
Nacional (UTN) for partial support through Grant PID Number SIUTNBA0006595.

\clearpage
\appendix

\section{\label{sec:global_complementarias}Global lockdown with complementary 
health policies}

\begin{figure*}[!ht]
\centering
\subfloat[Infected\label{fig:plot_infec_global_var_beta}]{
\includegraphics[scale=0.8]{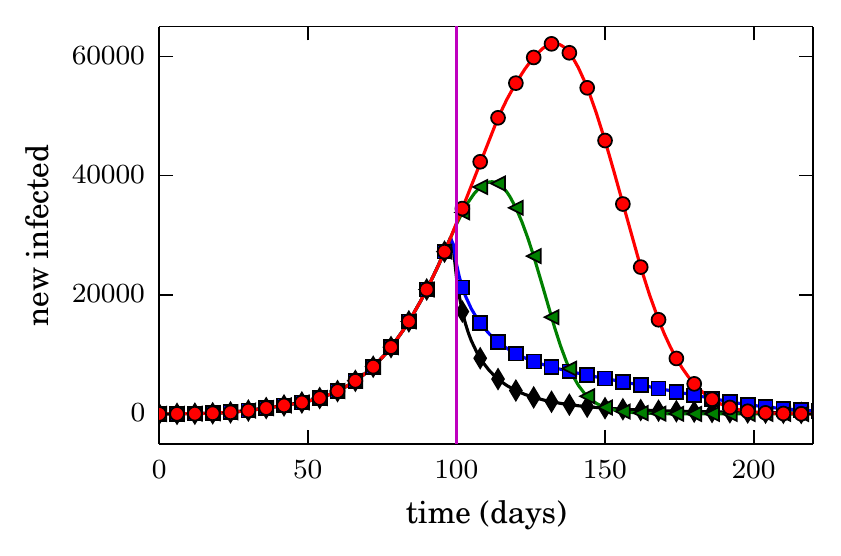}
}
\subfloat[Removed\label{fig:plot_recup_global_var_beta}]{
\includegraphics[scale=0.8]{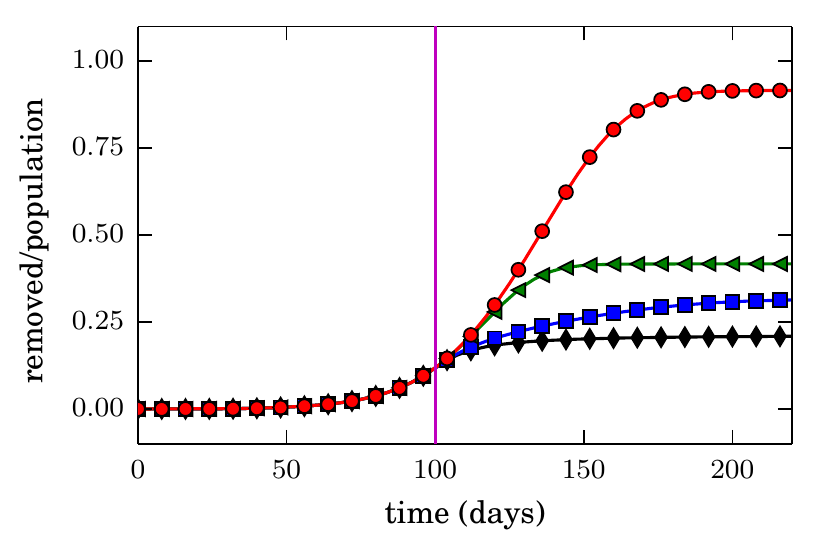}
}
\caption{\label{fig:infect_recup_global_betas} (a) Number of new infected 
people and (b) normalized number of removed individuals as a function of 
time. The plot in (b) is normalized with respect to the total population of the 
city ($4.32\,$M). The global lockdown started at the day 100 after the first 
infected was detected. This is indicated by a vertical line in (a) and 
(b). Before this day, the infection rate equals to $\beta_0=0.75$. However, 
from this day, the infection rate equals: \mydiamond{black} $(1/3)\beta_0$, 
\protect\tikz\protect\draw[black,fill=blue] (0,0) rectangle (0.3cm,0.3cm); 
$(1/2)\beta_0$ and \mytriangleleft{green}~$\beta_0$. 
\protect\tikz\protect\draw[black,fill=red] (0,0) 
circle (.9ex); corresponds to the scenario of no lockdown at all. In the last 
case, the infection rate remains constant along the simulation process and 
equals to $\beta_0=0.75$.} 
\end{figure*}

We analyze here the effects of complementary health policies during the global 
lockdown. Fig.~\ref{fig:plot_infec_global_var_beta} shows the number of new 
infected people along time for three different infection rate (see caption for 
details). It is also shown the evolution for the case of no lockdown at all. The 
mobility cutoff prevents the infection curves 
in Fig.~\ref{fig:plot_infec_global_var_beta} from growing almost immediately 
after the beginning of the lockdown. In this sense, the more additional health 
policies, the faster decrease of the new infected individuals. \\

Fig.~\ref{fig:plot_recup_global_var_beta} exhibits the number of individuals 
that passed over the disease as a function of time (see caption for details). 
These correspond to those individuals that previously appear as infected in 
Fig.~\ref{fig:plot_infec_global_var_beta}. First, it can be seen that the 
number of removed individuals increase since the outbreak of the disease. 
Second, it reaches a plateau soon after the lockdown is established. The plateau 
level, however, strongly depends on the complementary health policies. The 
more ``intense'' health policies (\textit{i.e.} the lower infection rate), the 
lower number of removed individuals. \\

\section{\label{sec:comp_local_global}Comparison between global and local 
lockdown}

This appendix compare the global and local lockdown in terms of 
the new infected individuals. Recall that the global lockdown ``cutoff'' the 
human mobility around the city. In this sense, all blocks are isolated from the 
rest regardless of whether they are ``infected'' or not. On the contrary, the 
mobility is suppressed (depending the infected threshold) only for the 
inhabitants of ``infected blocks'' in the local lockdown. \\

\begin{figure}[!ht]
\centering
\includegraphics[width=0.6\columnwidth]{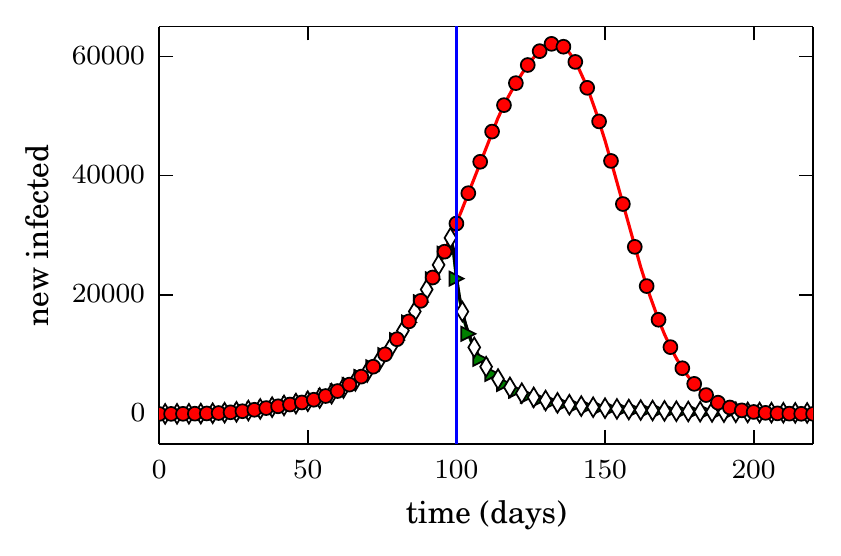}
\caption{\label{fig:infec_comp_local_global_rl_1} Comparison between the number 
new infected in the global (\mydiamond{white}) and local 
(\mytriangleright{green}) lockdown scenario. The global and local lockdown 
started at the day 100 after the first infected was detected. This is indicated 
by a vertical line. From this day, all those infected blocks in 
which the number of infected exceeds 0 are isolated (in the local lockdown 
scenario). \protect\tikz\protect\draw[black,fill=red] (0,0) 
circle (.9ex); corresponds to the scenario of no lockdown at all. From the 
lockdown implementation day, the infection rate switches from $\beta_0=0.75$ to 
$(1/3)\beta_0$ (in both cases).}
\end{figure}

Fig.~\ref{fig:infec_comp_local_global_rl_1} shows the number of new infected 
people along time implementing a global and local lockdown (see caption for 
details). It is also shown the evolution for the case of no lockdown at all. 
It can be seen that both curves matches along time. This means that it is not 
necessary isolated all blocks to reduce the infection. In this sense, as can be 
expected, only is necessary isolated those infected blocks. Therefore, we can 
conclude that the local lockdown is the "efficient" case of the global 
lockdown. \\

\section{\label{sec:testing_tracing}Implementation of the testing $\&$ 
back-tracing strategy}

We represent in this Section a possible testing procedure by means of a 
probability $p$ as follows:

\begin{enumerate}
 \item We first chose (and test) a block at random with probability $p$.
 \item If the chosen block is infected, we lock down the block until the 
disease disappears. If not, the block remains ``open''. 
 \item We trace back and test \textit{all} the individuals who visited the 
infected block.
 \item We lock down any of the above blocks if infected.
 \item We repeat this procedure every day.
\end{enumerate}

\begin{figure}[!ht]
\centering
\includegraphics[width=0.7\columnwidth]{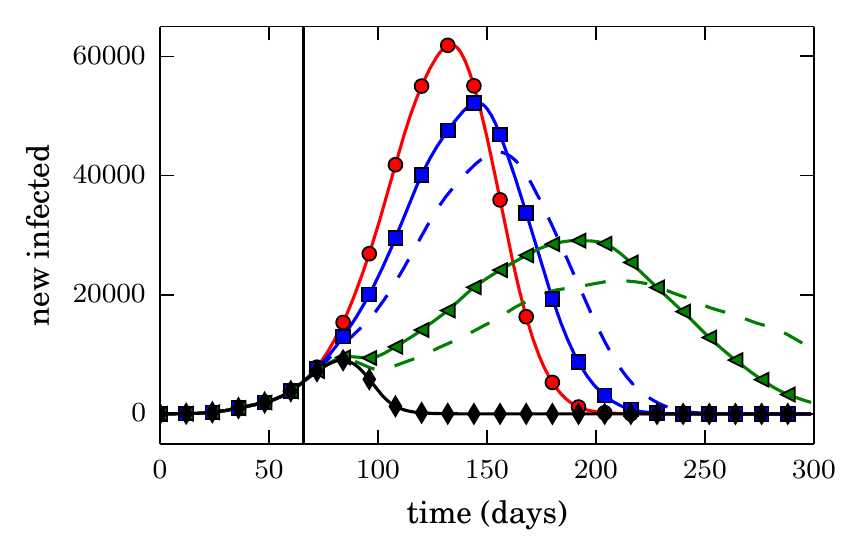}
\caption{\label{fig:plot_infec_rastreo} Number of new infected people as a 
function of time. The local lockdown is applied when the number of new 
infected equals to 5k (indicated by the vertical line). From the lockdown 
implementation day, we test each block with probability equal to: 
\mydiamond{black}~$100\%$, \mytriangleleft{green}~$75\%$ 
and \protect\tikz\protect\draw[black,fill=blue] (0,0) 
rectangle (0.3cm,0.3cm); $25\%$ (see text for details). 
\protect\tikz\protect\draw[black,fill=red] (0,0) circle (.9ex); corresponds to 
the scenario of no lockdown at all (\textit{i.e.} $p=0\%$). Dashed 
lines corresponds to the scenario including back tracing of the infected 
individuals. Those infected blocks are isolated until the infection inside the 
corresponding block disappears. We consider the isolation of the infected blocks 
as the only heath policy. Thus, the infection rate equals to $\beta_0=0.75$ all 
along the propagation process.}
\end{figure}

Fig.~\ref{fig:plot_infec_rastreo} shows the number of new infected people along 
time for three different testing probabilities (see caption for details). It is 
also shown the time evolution for the case of no lockdown at all. Notice that 
the propagation decreases dramatically for the ideal tested scenario. But the 
possibility of stopping the outbreak vanishes in a bad (or poor) testing 
scenario. As can be seen in Fig.~\ref{fig:plot_infec_rastreo} the back-tracing 
of the infected improves the mitigation strategy.\\

\begin{figure}[!ht]
\centering
\includegraphics[width=0.7\columnwidth]{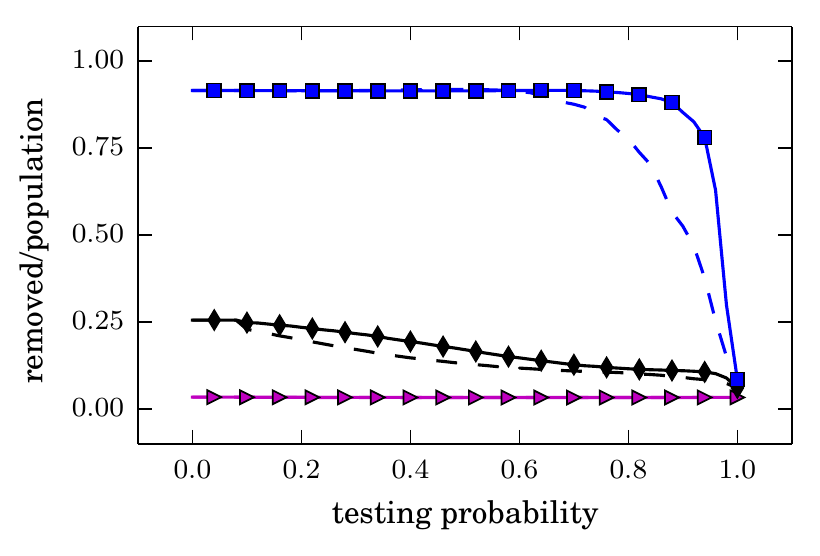}
\caption{\label{fig:plot_recup_rastreo} Normalized number of removed 
individuals at the end of the epidemic as a function of the testing 
probability. The normalization was done with respect to the city 
population. The local lockdown is applied when the number of new infected 
equals to 5k. From the lockdown implementation day, the infection rate switches 
from $\beta_0=0.75$ to: \protect\tikz\protect\draw[black,fill=blue] (0,0) 
rectangle (0.3cm,0.3cm);~$\beta_0$, \mydiamond{black}~$(1/2)\beta_0$ 
and \mytriangleright{magenta}~$(1/3)\beta_0$ (see text for more details). 
Dashed lines corresponds to the scenario including back tracing of the 
infected individuals. Those infected blocks were isolated until the infection 
within the corresponding block disappears.}
\end{figure}

Fig.~\ref{fig:plot_recup_rastreo} shows the number of removed individuals at 
the end of the epidemic in terms of the testing probability $p$ for three 
different complementary health policies (see caption for details). It can be 
seen that, in a scenario without complementary health policies (blue squares in 
Fig.~\ref{fig:plot_recup_rastreo}), a massive testing (of more than 
$80\%$) is the only tool to avoid a wide spreading on the population. Note 
that this percentage is significantly reduced to $60\%$ if a back tracing 
policy is also implemented.\\

Notice from Fig.~\ref{fig:plot_recup_rastreo} that the implementation of strict 
complementary health policies (such as the use of a mask and social distancing) 
improves the mitigation of the disease.\\

\section*{References}

\bibliographystyle{elsarticle-num}
\renewcommand{\bibname}{}
 \bibliography{paper}

\end{document}